\documentclass[11pt,a4paper]{article}
\usepackage{jheppub}

\usepackage{caption} 
\usepackage{subcaption} 
\usepackage{float} 
\usepackage{setspace} 
\usepackage{pgf}

\usepackage{multicol}

\usepackage{breqn}
\numberwithin{equation}{section}
\usepackage[T1]{fontenc} 
\usepackage[utf8]{inputenc}       

\usepackage[english]{babel}           

\usepackage{tabularx} 
\usepackage{adjustbox}
\usepackage{psfrag} 
\usepackage{sistyle} 

\usepackage{eurosym} 

\usepackage{epstopdf}

\usepackage{fancyhdr} 

\usepackage{slashed} 

\usepackage{fancybox}

\usepackage{setspace}
\usepackage{lscape}

\usepackage{newunicodechar}
\newunicodechar{ﬁ}{fi}
\newunicodechar{ﬀ}{ff}

\usepackage{tikz}
\usetikzlibrary{shapes}
\usepackage{verbatim}
\usetikzlibrary{positioning}

\usetikzlibrary{arrows,decorations.markings}

\title{F-theory and Heterotic Duality, Weierstrass Models from Wilson lines}

\author{Lilian Chabrol}

\affiliation{Institut de Physique Th\'eorique, 
Universit\'e Paris Saclay, CEA, CNRS\\
Orme des Merisiers \\
91191 Gif-sur-Yvette Cedex, France}
\emailAdd{lilian.chabrol@ipht.fr}
\abstract{We present how to construct elliptically fibered K3 surfaces via Weierstrass models which can be parametrized in terms of Wilson lines in the dual heterotic string theory. We work with a subset of reflexive polyhedras that admit two fibers whose moduli spaces contain the ones of the $E_{8}\times E_{8}$ or $\frac{Spin(32)}{\mathbb{Z}_{2}}$ heterotic theory compactified on a two torus without Wilson lines. One can then interpret the additional moduli as a particular Wilson line content in the heterotic strings. A convenient way to find such polytopes is to use graphs of polytopes where links are related to inclusion relations of moduli spaces of different fibers. We are then able to map monomials in the defining equations of particular K3 surfaces to Wilson line moduli in the dual theories. Graphs were constructed developing three different programs which give the gauge group for a generic point in the moduli space, the Weierstrass model as well as basic enhancements of the gauge group obtained by sending coefficients of the hypersurface equation defining the K3 surface to zero.}


\arxivnumber{}

\begin{document}
\maketitle
\flushbottom
\renewcommand{\arraystretch}{1.5}

\section{Introduction}

F-theory compactified on elliptically fibered K3 surfaces is believed to be dual at the quantum level to the heterotic string compactified on a two torus with Wilson lines \cite{vafa_evidence_1996,morrison_compactifications_1996-1,morrison_compactifications_1996,berglund_heterotic_1998}. In particular one should be able to relate the complex parameters of the moduli space on the F-theory side to the ones on the heterotic one as their moduli space are the same, namely the Narain space \cite{narain_new_1986,aspinwall_k3_1996} 
	\begin{equation}
	\label{narain}
	\mathcal{M} = \left( O(2)\times O(18) \right) \backslash O(2,18) \slash O(2,18,\mathbb{Z}).
\end{equation}
	The study of the full moduli space is however a tedious exercise and one wants to focus on subspaces with fewer complex modular parameters. In heterotic string one can consider for example compactifications on a two torus with Wilson lines parametrized by few moduli. In F-theory, one can choose an algebraic K3 with a large Picard number $p$, as its modular space is parametrized by $20-p$ complex variables \cite{schuett_elliptic_2009}. Now, a particularly interesting way to construct K3 surfaces is to use lattice polarized K3 obtained by considering reflexive polyhedras in 3 dimensions which define hypersurfaces in toric varieties \cite{laza_arithmetic_2013,laza_calabi-yau_2015-1}. Thanks to Kreuzer and Skarke \cite{kreuzer_classification_1998} it is possible to have access to the totality of the 4319 different reflexive polyhedras in 3 dimensions. One can then first construct K3 surfaces, and afterwards elliptically fibered K3s by considering a particular subdivision of the fan in the dual polytope dictated by a choice of a two dimensional reflexive subpolytope which plays the role of the fiber. In particular it is possible to classify the different reflexive polyhedras with respect to their Picard number $p$, and therefore focus on moduli spaces associated to particular K3 surfaces with a low number of moduli.
	
	 The duality between F-theory and heterotic string has been written explicitly for only two of the 4319 different K3 surfaces one can construct via reflexive polytopes. First the duality between the parameter of a Weierstrass model presenting a particular $E_{8}\times E_{8}$ singularity and the complex structure and Kahler moduli of the two torus on which the $E_{8}\times E_{8}$ heterotic string is compactified was constructed in \cite{cardoso_duality_1996}. Later it was found that a particular reflexive polyhedra admitting two fibrations has for gauge groups $E_{8}\times E_{8}$ and $\frac{Spin(32)}{\mathbb{Z}_{2}}$ \cite{candelas_f-theory_1997}. In a more general case with three moduli, Malmendier and Morrison showed that a particular polytope with again two fibers with gauge group $E_{7}\times E_{8}$ and $\frac{Spin(28)\times SU(2)}{\mathbb{Z}_{2}}$ is related to compactifications of heterotic strings with one Wilson line moduli \cite{malmendier_k3_2015}. All of this suggests that compactifying F-theory on elliptically fibered K3 surfaces described by polytopes with two fibers seem to be related in some cases to the compactification of both heterotic strings with Wilson line moduli.
	 
	  Here we show that if we focus on particular reflexive polyhedras that are linked in some way to the $E_{8}\times E_{8}/\frac{Spin(32)}{\mathbb{Z}_{2}}$ polytope, we can understand the Wilson line structure of the dual heterotic strings. This is due to the fact that we can recover the torus on which we compactify the heterotic strings as a particular subspace of the moduli spaces of the elliptically fibered K3s. To find these polytopes we construct graphs where a link between two polytopes $M^{+}$ and $M^{-}$ is drawn if, for every elliptically fibered K3 surface obtained via $M^{+}$, there exist a limit in the moduli space where one obtains elliptically fibered K3 surfaces of the other polytope $M^{-}$. In particular we will consider the limit where we send monomials of the hypersurface equation defining the K3 surface associated to $M^{+}$ to zero, which is equivalent to removing a point in $M^{+}$. This can be seen as an extension of the notion of chains presented by Kreuzer and Skarke in \cite{kreuzer_classification_1998}. Focusing on polytopes which have two fibers, links between polytopes then correspond to inclusion relations between the moduli spaces of elliptically fibered K3s. Considering polytopes which are linked to $E_{8}\times E_{8}/\frac{\text{Spin}(32)}{\mathbb{Z}_{2}}$, we show that additional monomials in the hypersurface equation which defines the elliptically fibered K3s on which we compactify on correspond to additional Wilson line moduli in both the $E_{8}\times E_{8}$ and $\frac{Spin(32)}{\mathbb{Z}_{2}}$ heterotic strings. Using this Wilson line/monomial duality we can construct Weierstrass models of elliptically fibered K3s which are not directly obtained from reflexive polyhedras. They can then be interpreted as a certain Wilson line content in the dual heterotic theories. Finally, we show that in some cases this notion of \textit{Wilson line description of K3 surfaces} can be extended to polytopes with more than two fibers. This should be helpful to explicitly understand the duality between F-theory compactified on K3s and heterotic string on a two torus, and eventually in compactifications to lower dimensions involving K3 surfaces. \\

	The paper is organised as follows: in section \ref{section2} we present some basic properties of reflexive polyhedras and how they define elliptically fibered K3 surfaces. In section \ref{section3}, we present several computer programs which are helpful for constructing graphs of polytopes. They were written on SageMath and with the help of the package PALP \cite{the_sage_developers_sagemath_2019,braun_palp_2012,kreuzer_palp_2004}.  The first program uses the extended Dynkin diagram structure of reflexive polyhedras with fibers in order to construct tables of gauge groups for each fibration of every reflexive polytope. The second program gives the corresponding Weierstrass model for every fiber of reflexive polytopes. The third one uses this Weierstrass model and finds the enhancements one can obtain by simply sending the coefficient which parametrize the hypersurface equation of the K3 in some toric varieties to zero. This can be particularly useful to construct graphs of polytopes and we show how one can link polytopes up to three moduli. In the appendix we present typical outputs of the programs and explain how to use them. We are making the computer programs available on GitHub at \textit{https://github.com/lilianChabrol/ReflexivePolyhedras}.
To summarize, here are the three SageMath programs available online
	\begin{itemize}
		\item Program 1 (Typical output in Appendix A): Gauge groups from the extended Dynkin diagram structure in the $N$ lattice.
		\item Program 2 (Typical output in  Appendix B): Determination of the Weierstass model of the corresponding elliptically fibered K3.
		\item Program 3 (Typical output in Appendix C): Possible enhancements of the gauge groups for each fibers by sending defining coefficients of the hypersurface to zero.
	\end{itemize}
	Finally in section \ref{section4} we present a \textit{Wilson line description of K3 surfaces} by considering a particular graph of polytope which goes up to 6 moduli, or equivalently in this case four Wilson line moduli on the heterotic side. We then show how to construct Weierstrass models of elliptically fibered K3s which one can directly interpret in the dual theory as a particular Wilson line content.

\section{Reflexive Polyhedras and Elliptically Fibered K3s}
\label{section2}
\subsection{Hypersurfaces on Projective Spaces and Fano Variety}
Here we introduce various notations about reflexive polyhedras and present briefly results about toric Fano varieties. Detailed constructions of toric Fano varieties have been widely discussed in the litterature (see e.g. \cite{batyrev_dual_1993,cox_toric_nodate}). A pedagocical introduction to toric geometry can be found in \cite{skarke_string_1999}.

Let us consider two dual lattices $M$ (Monomials) and $N$ (faN) in $\mathbb{Z}^{n}$ with real extension $M_{\mathbb{R}}$ and $N_{\mathbb{R}}$ and an associated product $<*,*>: M\times N \rightarrow \mathbb{Z}$. We note $\Delta$ an integral convex polytope whose vertices are in $M$ and which contains only the origin as an interior point. We then define the dual of $\Delta$ as 
\begin{equation}
	\bigtriangledown \equiv \left\lbrace v \in N_{\mathbb{R}}: <w,v> \ \geq -1 \text{ for all }w \in \Delta \right\rbrace.
\end{equation}
As usual we consider $\Delta$ to be reflexive, meaning that $\bigtriangledown$ is also convex, only contains the origin and has its vertices $v_{i}$ in $N$.
 The normal fan of the polytope $\Delta$ whose rays are the vertex of $\bigtriangledown$ then defines a projective toric variety $P_{\Delta}$ (which is Fano if and only if $\Delta$ is reflexive, which it is in this paper).

Now that we constructed this Fano toric variety using the pair of polytopes ($\Delta$, $\bigtriangledown$) it is possible when $n=3$ to construct 
K3 surfaces as hypersurfaces in $P_{\Delta}$. Explicitly, one associates a variable $x_{i}$ to each of the vertices $v_{i}$ of the polytope $\bigtriangledown$ in $N$. The K3 surface $X_{\Delta}$ can then be written as the locus in  $P_{\Delta}$ of

\begin{equation}
	\label{hypersurfaceEquation}
	\sum_{m \in \Delta \cap M} c_{m} \prod_{k = 1}^{i} x_{k}^{<m,v_{k}>+1}=0
\end{equation}
with $c_{m}\in \mathbb{C}$.

We can then construct in some cases an elliptically fibered K3 as $X_{\Delta}$ together with a surjective morphism $\pi: X_{\Delta} \rightarrow \mathbb{P}_{1}$ such that generic fiber are genus one elliptic curves. They can be constructed by considering the K3 surface we just described, as well as finding a subpolytope $\bigtriangledown^{(2)}$ of $\bigtriangledown$ in the $N$ lattice which plays the role of the fiber of the elliptic K3\footnote{Such subpolytope cannot be found sometimes, in particular for small Picard numbers.}. There are 16 reflexive polyhedras for $n=2$, and 4319 reflexive polytopes for $n=3$ \cite{kreuzer_classification_1998}. It is then possible to obtain Weierstrass models of elliptically fibered K3 surfaces upon a choice of a fan which contains as rays points of the fiber $\bigtriangledown^{(2)}$.
Quite amazingly, and upon a particular choice of a fan which will be described in section \ref{choiceVector}, the gauge groups associated to singularities of the elliptically fibered K3s can be read off directly once one chooses a particular subpolytope $\bigtriangledown^{(2)}$ \cite{candelas_duality_1998}. Indeed, Candelas and Font noticed that the points located on both sides of the fiber of the polytope $\bigtriangledown$ in the $N$ lattice are exactly the extended Dynkin diagrams which correspond to the gauge groups associated to singularities appearing in the Weierstrass model via the Kodaira and Néron classification \cite{kodaira_compact_1963,neron_modeles_1964} (see Figure \ref{extendedDynkin}). This was later explained by Perevalov and Skarke in \cite{perevalov_enhanced_1997}. Depending on which of the 16 two dimensional reflexive polyhedras is the fiber, additional contribution coming from the Mordell-Weil group of rational sections of the elliptic fibration can occur \cite{mayrhofer_mordell-weil_2014,braun_geometric_2013,klevers_f-theory_2015,cvetic_discrete_2017,cvetic_tasi_2018}. In particular the fibers F1, F2 or F4 give additional discrete symmetries $\mathbb{Z}_{\#}$ and fibers F13, F15 and F16 quotient by discrete symmetries $\frac{1}{\mathbb{Z}_{\#}}$\footnote{Using the notations of \cite{klevers_f-theory_2015}. The polytopes are i1,i2,i4 and i9,i7,i6 using notations of \cite{grassi_weierstrass_2012}.}. Finally, additional contribution of $U(1)$s or $SU(\#)$ factors can appear, depending on how the polytope $\bigtriangledown^{(2)}$ intersects with $\bigtriangledown$.
\subsection{Invariant Parameters of the Moduli Space}
\label{sectionInvariant}
The number of complex moduli for a K3 surface with Picard number $p$ is $20-p$. Previously we defined an algebraic K3 as an hypersurface \eqref{hypersurfaceEquation} in the toric variety $P_{\Delta}$ whose number of parameters is a priori given by the number of points in $\Delta \cap M$ if we consider \eqref{hypersurfaceEquation}. However different sets of those parameters correspond to the same point in the moduli space. For example several of the coefficients can be put to 1 by a reparametrization of the coordinates in the projective space. In order to properly define complex parameters on the moduli space of the K3 surface we use the construction developed in \cite{candelas_type_2015}. It was shown there that monomials defined by points interior to facets in $\Delta \cap M$ can be removed by an appropriate change of coordinates for the different reflexive polyhedras they considered. We therefore restrict the hypersurface equation \eqref{hypersurfaceEquation} to the integral points $m \in \text{Edges}\left(\Delta \cap M\right) \equiv \text{Edg}(\Delta)$ as well as the origin. The hypersurface equation can then be written as
\begin{equation}
 \label{hypersurfaceEquation2}
 	H = -c_{0} \prod_{k = 1}^{n} x_{k} + \sum_{m \in \text{Edg}(\Delta)} c_{m} \prod_{k = 1}^{n} x_{k}^{<m,v_{k}>+1}=0
\end{equation}
with $v_{k}$ rays of the normal fan $P_{\Delta}$.
Due to the strong link between the period map of K3 surfaces and their moduli spaces \cite{schuett_elliptic_2009}, one can seek for parameters of the moduli space by considering the fundamental period of the holomorphic two-forms which can be written in our case as \cite{berglund_periods_1994}
\begin{equation}
	\bar{w}_{00} = -\frac{c_{0}}{(2\pi i)^{n}}\oint_{\mathcal{C}} \frac{\text{d}x_{1}\wedge ... \wedge \text{d}x_{n}}{H}
\end{equation}
with $\mathcal{C}$ a product of cycles that enclose the hypersurface defined by $x_{i} = 0$ \cite{candelas_type_2015}. This can be recast as
\begin{equation}
	\label{cycle2}
	\bar{w}_{00} = \frac{1}{(2\pi i)^{n}}\oint_{\mathcal{C}} \frac{\text{d}x_{1}\wedge ... \wedge \text{d}x_{n}}{\prod_{k=1}^{n}x_{k}}\sum_{l=0}^{\infty} \tilde{H}^{l}
\end{equation}	
	with
	\begin{equation}
		\label{devf}
		\tilde{H} = \sum_{m \in \text{Edg}(\Delta)}\frac{ c_{m} \prod_{k = 1}^{n} x_{k}^{<m,v_{k}>+1}}{c_{0}\prod_{k=1}^{n}x_{k}}.
	\end{equation}
	 The only non zero terms in \eqref{cycle2} are the constant terms\footnote{By the residue theorem.} in the development of $\tilde{H}^{l}$. The fundamental period of the holomorphic two-forms can therefore be parametrized by the following invariants
	\begin{equation}	
	\text{Moduli} \sim \frac{l!}{c_{0}^{l}}\prod_{m \in \text{Edg}(\Delta)}	\frac{c_{m}^{l_{m}}}{l_{m}!}
	\end{equation}
	such that 
	\begin{equation}
		\label{conditions}
		\sum_{m \in \text{Edg}(\Delta)} l_{m} = l \quad \text{and} \quad \forall k \sum_{m \in \text{Edg}(\Delta)} l_{m}\left(<m,v_{k}>+1\right) = l.
	\end{equation}
	Taking the second equation one can then simply look for inequivalent linear relations in the $M$ lattice such that $\sum l_{m}\cdot m  = (0,0,0)$ with $l_{m}$s positive and minimal. By a change of variables of these invariants one can in fact look for inequivalent linear relations between points in the edges of $\Delta$ such that \eqref{conditions} is verified but this time with $l_{m}$ in $\mathbb{Z}$ and $|l_{m}|$ minimal. The complex parameters can then be taken to have the following form
	\begin{equation}
	\label{moduli}
		\text{Moduli}\sim \left(\prod_{m \in \text{Edg}(\Delta)} c_{m}^{l_{m}}\right) c_{0}^{-l}.
	\end{equation}
	As an example let us take the polytope $M476$, with Picard number equal to 16 i.e. 4 moduli. Its vertices are given by
	\begin{equation}
	M476: 
\underbrace{\left(1,\,0,\,0\right)}_{\ \ (1)_{M}}, \underbrace{\left(0,\,1,\,0\right)}_{\ \ (2)_{M}},
\underbrace{\left(0,\,0,\,1\right)}_{\ \ (3)_{M}}, \underbrace{\left(-4,\,-2,\,-1\right)}_{\ \ (4)_{M}},
\underbrace{\left(-5,\,-3,\,-1\right)}_{\ \ (5)_{M}}, \underbrace{\left(-1,\,-1,\,1\right)}_{\ \ (6)_{M}}.
	\end{equation}
	An additional point, $(7)_{M} = \left(-3,-2,0\right)$, is situated on the edges of the polytope. We can thus consider four inequivalent linear relations between these points, a possibility being
	\begin{align}
		(5)_{M}+(6)_{M}-2\cdot(7)_{M} \quad ,& \quad (7)_{M}+2\cdot(2)_{M}+3\cdot(1)_{M} \\ (3)_{M}-(1_{M})-(2)_{M}-(6)_{M} \quad ,& \quad (4)_{M}-(1)_{M}-(2)_{M}-(5)_{M}
	\end{align}
	which leads using \eqref{moduli} to the complex parameters of the moduli space
	\begin{equation}
		M476 \text{ (4 Moduli): } \quad	 \frac{c_{5}c_{6}}{c_{7}^2} \quad , \quad  \frac{c_{7}c_{2}^{2}c_{1}^{3}}{c_{0}^{6}}\quad , \quad \frac{c_{3}c_{0}^2}{c_{1}c_{2}c_{6}} \quad , \quad \frac{c_{4}c_{0}^2}{c_{1}c_{2}c_{5}}.
	\end{equation}
\section{Obtaining Data on Elliptically Fibered K3s}
\label{section3}
	We now present three computer programs that allow to obtain different information about elliptically fibered K3 surfaces automatically. We note $M\#$ the polytope $\Delta$ in the $M$ lattice corresponding to $ReflexivePolytope(3,\#)$ in SageMath\footnote{The vertices of each of the polytopes presented in this paper are written in the Appendix \ref{appendixD}.}. We write schematically what is done in Sagemath, more accurate descriptions of the computer programs as well as the programs themselves are available on GitHub at
	\begin{center} \textit{https://github.com/lilianChabrol/ReflexivePolyhedras}.
	\end{center}
	\subsection{Extended Dynkin Diagram from Polyhedras}

	As discussed in the first part of this paper it is possible to have access to the gauge structure of an elliptically fibered K3 upon a choice of reflexive polytope $(\Delta,\bigtriangledown)$, and a choice of fiber $\bigtriangledown^{(2)}$.  We now present a generic way to find the gauge group associated to each fiber of every reflexive polytope in the Kreuzer-Skarke classification of reflexive polyhedra in 3 dimensions. 
		
\begin{center}
	\begin{figure}[h!]
	\caption{$E_{7}\times E_{7}$ and $\frac{SO(24)\times SU(2)^{2}}{\mathbb{Z}_{2}}$ fiber of the polytope $M476$. The points in blue draw the extended Dynkin diagram of $E_{7}$s on the left, $SO(24)$ on the right. The contribution of $SU(2)$s from the fiber are symbolised by red points. The fiber being F13 there is an additional contribution of $\frac{1}{\mathbb{Z}_{2}}$.}
	\label{extendedDynkin}
	\centering
		\includegraphics[scale=0.18]{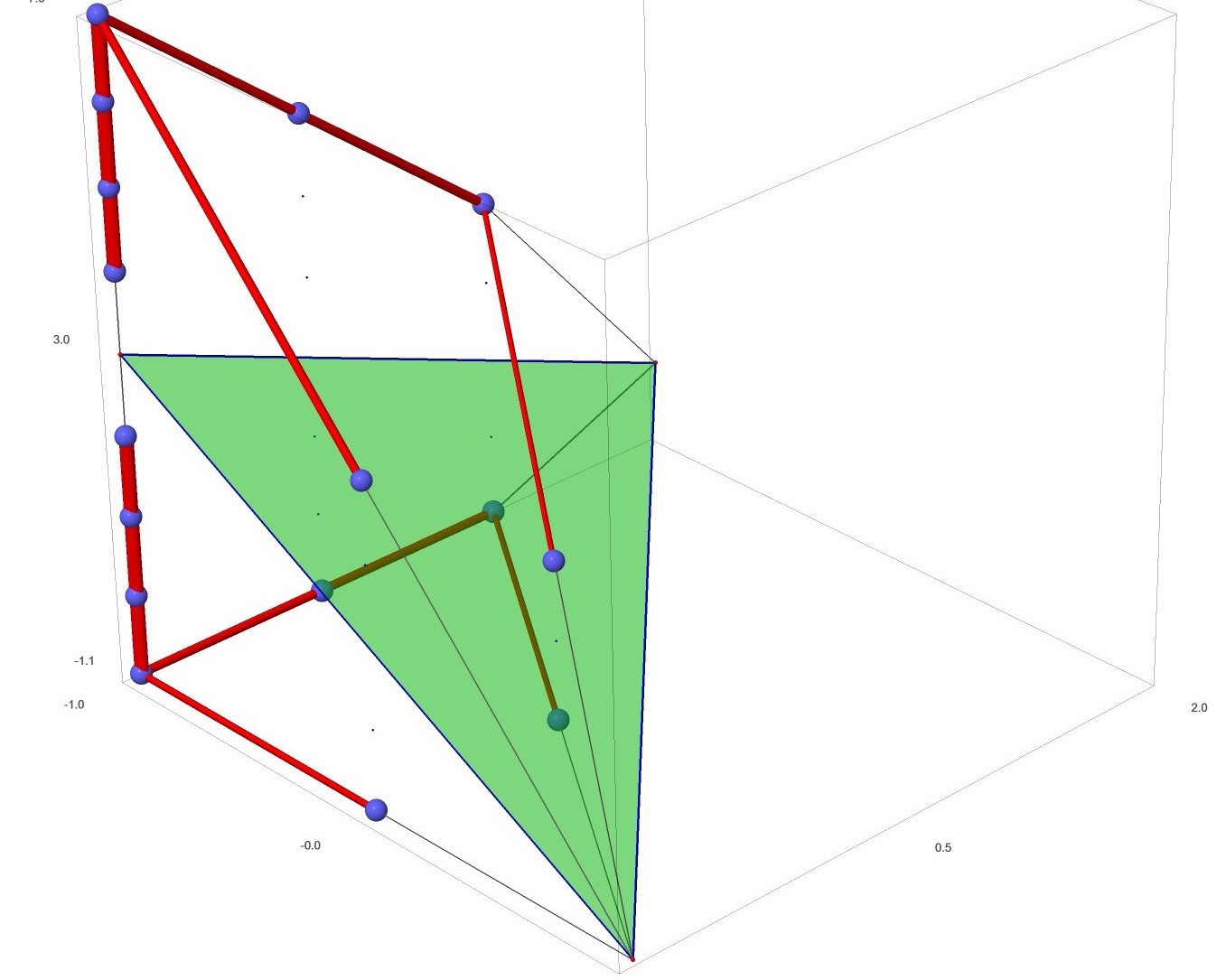}
		\hspace{1cm}
		\includegraphics[scale=0.18]{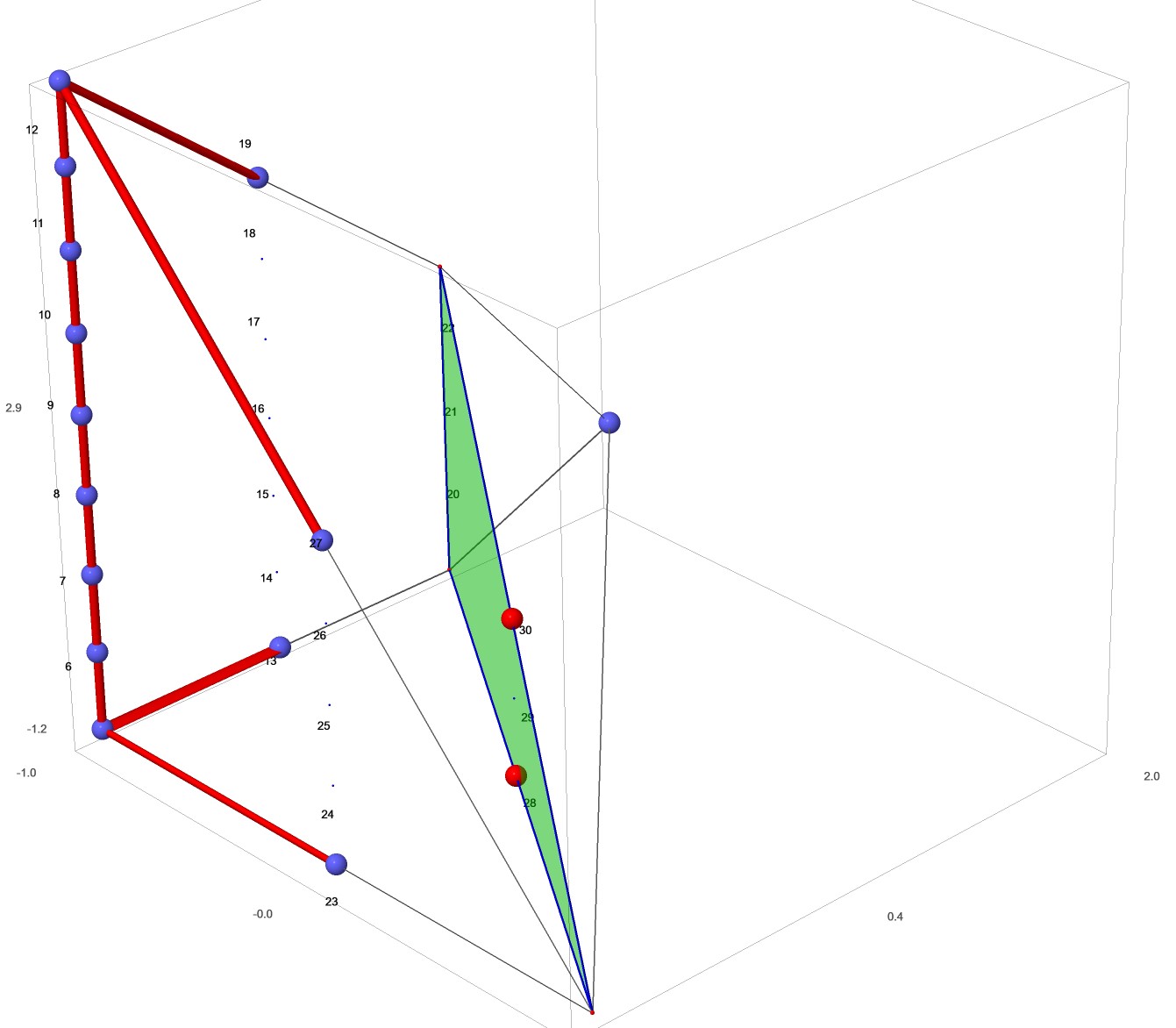}
	\end{figure}
	\end{center}

		We first find all two dimensional reflexive polyhedras $\bigtriangledown^{(2)}$ which are subpolytopes of $\bigtriangledown$ modulo $SL(3, \mathbb{Z})$ transformations in the $N$ lattice. Then we identify which of the 16 possible two dimensional reflexive polytope corresponds to each of the fibers $\bigtriangledown^{(2)}$. This permits in particular to know if the fiber contains product or quotient by discrete symmetry group  \cite{klevers_f-theory_2015}: F1, F2 and F3 contribute to a product by $\mathbb{Z}_{3}$, $\mathbb{Z}_{2}$ and $\mathbb{Z}_{4}$ respectively while fibers F13, F15 contribute by  $\frac{1}{\mathbb{Z}_{2}}$ and F16 by $\frac{1}{\mathbb{Z}_{3}}$. We do not write the additional contributions of $U(1)$ factors coming from the Mordell Weil group as in the end the gauge group must be of rank 18. We however look for additional $SU(\#)$ contribution from the fiber: if polytopes $\bigtriangledown^{(2)}$ and $\bigtriangledown$ have a common edge with $n$ points, then there appears an additional $SU(n-1)$ part in the final gauge group\footnote{Equivalently the rank of the $SU(\#)$ can be seen by looking at the number of interior points in the common edge. See the red points in Figure \ref{extendedDynkin}.}.  Finally, the fiber $\bigtriangledown^{(2)}$ dividing $\bigtriangledown$ into two parts,   we look at points "above" and "below" the fiber and read off the extended Dynkin diagrams. \\
		
		As an example let us consider the polytope $M476$. In Figure \ref{extendedDynkin} we represent the dual polytope $N476$ of $M476$ for the two inequivalent fibrations $\bigtriangledown^{(2)}$ it contains.  On the left one can read off two extended Dynkin diagram of $E_{7}$. On the right there is a $SO(24)$ as well as $\frac{1}{\mathbb{Z}^{2}}$ comming from the fiber $F13$, and $SU(2)\times SU(2)$ contribution due to the intersection of $\bigtriangledown^{(2)}$ and $\bigtriangledown$ symbolised by red points.
	The results for K3 surfaces with Picard number 19 and 18 i.e. one and two complex parameters respectively, are presented in the Tables \ref{pic19} and \ref{pic18}. They were compared with results of an unpublished paper \cite{font_comments_nodate-1}  presented at a seminar at CERN \cite{font_comments_2019}. The result with Picard 17 and 3 complex parameters is in the Appendix \ref{appendixPic} (Table \ref{picard17}). Tables with complex parameters up to 5 moduli are available on GitHub, and up to 10 moduli for elliptically fibered K3s admitting only two inequivalent fibrations.
\begin{table}[h!]
\caption{Gauge groups for polytopes with Picard 19. Columns represent the inequivalent fibers $\bigtriangledown^{(2)}$ dividing the dual $N\#$ of $M\#$ into two parts.}
\begin{center}
\begin{tabular}{|c|c|c|c|c|}
\hline
M0 & 
$\frac{SO(16) \times SO(16)}{Z_{2}}$  &
$\frac{SU(12) \times E_{6}}{Z_{3}}$  &
$E_{8} \times E_{8}$  &
$\frac{E_{7} \times E_{7}\times SU(4)}{Z_{2}}$ 
\\
\hline
M2 & 
$\frac{E_{7} \times SO(20)}{Z_{2}}$  &
$\frac{U1 \times SU(18)}{Z_{3}}$  &
$E_{8} \times E_{8}\times SU(2)$  &
\\
\hline
\end{tabular}
\label{pic19}
\end{center}
\end{table}
\begin{table}[h!]
\caption{Gauge groups for polytopes with Picard 18. Columns represent the inequivalent fibers $\bigtriangledown^{(2)}$ dividing the dual $N\#$ of $M\#$ into two parts.}
\begin{center}
\begin{adjustbox}{max width=\linewidth}
\begin{tabular}{|c|c|c|c|c|c|c|c|}
\hline
M3 & $SO(14) \times E_{7}$  &$SO(14) \times SU(9)$  &$\frac{SU(12) \times SO(8)}{Z_{2}}$  &$\frac{E_{6} \times E_{6}\times SU(3)SU(3)}{Z_{3}}$ &$E_{8} \times E_{8}\times Z_{3}$  & & \\ \hline
M4 & $E_{8} \times E_{8}\times Z_{3}$  &$E_{6} \times SO(14)\times SU(3)$  &$E_{7} \times E_{7}$  &$\frac{SU(10) \times SO(12)}{Z_{2}}$  &$\frac{SU(9) \times SU(9)}{Z_{3}}$  & & \\ \hline
M5 & $E_{7} \times E_{7}\times SU(2)$  &$SU(10) \times E_{6}$  &$\frac{SO(16) \times SO(12)\times SU(2)}{Z_{2}}$ &$\frac{E_{7} \times SO(12)\times SU(4)}{Z_{2}}$ &$E_{8} \times E_{7}$  &$\frac{SU(6) \times SU(12)}{Z_{3}}$  &\\ \hline
M6 & $E_{6} \times E_{7}\times SU(3)$  &$E_{7} \times E_{8}$  &$E_{8} \times E_{8}\times Z_{3}$  &$SO(14) \times SO(14)$  &$SO(10) \times SU(11)$  &$\frac{E_{6} \times SU(9)\times SU(3)}{Z_{3}}$ &$\frac{SU(8) \times SO(16)}{Z_{2}}$  \\ \hline
M7 & $E_{7} \times E_{8}$  &$\frac{SU(10) \times E_{7}}{Z_{2}}$  &$\frac{SU(3) \times SU(15)}{Z_{3}}$  &$E_{6} \times SO(18)$  & & &\\ \hline
M10 & $\frac{SO(16) \times SO(16)}{Z_{2}}$  &$\frac{E_{7} \times E_{7}\times SU(2)SU(2)}{Z_{2}}$ &$E_{8} \times E_{8}\times Z_{4}$  &$\frac{SU(16) \times U1}{Z_{2}}$   & & &\\ \hline
M11 & $\frac{SO(16) \times E_{7}\times SU(2)}{Z_{2}}$ &$E_{8} \times E_{8}\times Z_{4}$  &$E_{8} \times E_{7}\times SU(2)$  &$\frac{SO(12) \times SO(20)}{Z_{2}}$  &$SU(16) \times U1$  & & \\ \hline
M16 & $SO(18) \times E_{6}$  &$\frac{SU(15) \times U1\times SU(3)}{Z_{3}}$ &$E_{7} \times E_{8}$  &$\frac{SU(10) \times E_{7}}{Z_{2}}$  & & &\\ \hline
M88 & $E_{8} \times E_{8}$  &$\frac{SO(32) \times U1}{Z_{2}}$  & & & & &\\ \hline
\end{tabular}
\end{adjustbox}
\end{center}
\label{pic18}
\end{table}

	\subsection{Weierstrass Model, Gauge Groups and Basic Enhancements}
\label{choiceVector}
	The computer program introduced in section \ref{section3} is particularly interesting to determine the gauge group at a generic point in the moduli space associated to any fiber of any reflexive polytope in three dimensions. It would however be interesting to get the Weierstrass model which correspond to these gauge groups in order to find their enhancements for particular values of the moduli. Some of the enhancements can then be found quite easily by removing points in the polytope $\Delta$ in the $M$ lattice which amounts to sending to zero a coefficient in the hypersurface equation which defines the K3 surface. This is what the second and third program do: find the Weierstrass model, and the enhancements described above\footnote{See Appendix \ref{program2} and \ref{program3} for the output of the programs.}.
	
\begin{figure}[H]
			\begin{center}
				\begin{tikzpicture}
					\draw (0,0) node[scale=4,color=red] {.};
					\draw (-1,0) node[scale=4] {.};
					\draw (1,0) node[scale=4] {.};
					\draw (-2,1) node[scale=4] {.};
					\draw (-1,1) node[scale=4] {.};
					\draw (0,1) node[scale=4] {.};
					\draw (1,1) node[scale=4] {.};
					\draw (2,1) node[scale=4] {.};
					\draw (0,-1) node[scale=4] {.};
					
					\draw [-] (-1,1)--(0,-1);
					\draw [-] (-1,1)--(2,1);
					\draw [-] (2,1)--(0,-1);
					
					\draw (-1,1) node[scale=1,anchor=south] {$v_{1}$};
					\draw (2,1) node[scale=1,anchor=south] {$v_{2}$};	
					\draw (0,-1) node[scale=1,anchor=north] {$v_{3}$};

					\draw (5,0) node[scale=4,color=red] {.};
					\draw (5,1) node[scale=4] {.};
					\draw (5,-1) node[scale=4] {.};
					\draw (6,0)	 node[scale=4] {.};
					\draw (6,-1) node[scale=4] {.};
					\draw (6,1) node[scale=4] {.};
					\draw (4,0) node[scale=4] {.};
					\draw (4,-1) node[scale=4] {.};
					\draw (4,1) node[scale=4] {.};
					
					\draw (5,1) node[scale=1,anchor=south] {$v_{1}$};
					\draw (5,-1) node[scale=1,anchor=north] {$v_{2}$};
					\draw (4,0) node[scale=1,anchor=east] {$v_{3}$};
					\draw (6,0) node[scale=1,anchor=west] {$v_{4}$};
					
					\draw [-] (4,0)--(5,1);
					\draw [-] (4,0)--(5,-1);
					\draw [-] (6,0)--(5,1);
					\draw [-] (6,0)--(5,-1);
					
					\draw (9,0) node[scale=4,color=red] {.};
					\draw (9,-1) node[scale=4] {.};
					\draw (8,-1) node[scale=4] {.};
					\draw (8,0) node[scale=4] {.};
					\draw (8,1) node[scale=4] {.};
					\draw (8,2) node[scale=4] {.};
					\draw (10,-1) node[scale=4] {.};
					\draw (11,-1) node[scale=4] {.};
					\draw (10,0) node[scale=4] {.};
					\draw (9,1) node[scale=4] {.};
					
					\draw [-] (8,-1)--(9,1);
					\draw [-] (9,1)--(11,-1);
					\draw [-] (8,-1)--(11,-1);
					
					\draw (8,-1) node[scale=1,anchor=north] {$v_{1}$};
					\draw (9,1) node[scale=1,anchor=south] {$v_{2}$};
					\draw (11,-1) node[scale=1,anchor=west] {$v_{3}$};
				\end{tikzpicture}			
			\end{center}
			\caption{In order to obtain the groups associated to the extended Dynkin diagrams on the $N$ lattice we consider the following rays when the two dimensional subpolytopes are F13, F15 and F16.}
			\label{polytopes}
		\end{figure}
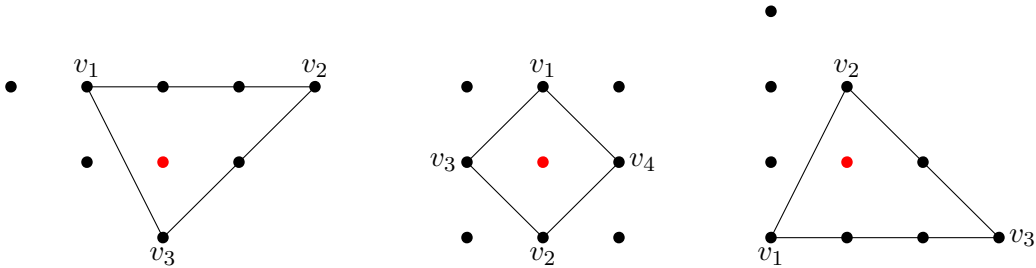

	We first look at the polytope $\bigtriangledown$ in the $N$ lattice. As explained in the introduction we then find inequivalent subpolytope $\bigtriangledown^{(2)}$ of dimension 2 in $\bigtriangledown$. For each of this 2 dimensional polytope we want to associate homogenous coordinates such that it describes the fiber. For most cases one can just associate one of them to each vertices of the subpolytope and obtain later the gauge groups expected from reading the extended Dynkin diagrams directly on the polytope $\bigtriangledown$. However in 3 cases (F13, F15 and F16) out of the 16 possible two dimensional reflexive polytopes, considering the vertices will not lead to these groups. This is due to the fact that for these particular polytopes the fibrations admit more than one section \cite{candelas_f-theory_1997}. Using a similar construction to the one of \cite{font_comments_2019} and in an upcoming paper \cite{font_comments_nodate-1}, we then consider the homogeneous coordinates $x_{i}$ of the fiber to be associated to the points as described in Figure \ref{polytopes}.
To define coordinates $(s,t)$ on the base space $\mathbb{P}^{(1)}$ we seek for two vectors $v_{s}$ and $v_{t}$, "above" and "below" the fiber. A fast way to obtain the appropriate Weierstrass model with correct ADE singularities, which correspond to the extended Dynkin diagrams seen in $\bigtriangledown$, is then to seek for the closest vectors to the fiber in $\bigtriangledown \cap N$. 
	
	Finally we write the hypersurface equation by considering the points on the edges of $\Delta$ and using equation \eqref{hypersurfaceEquation}. To each of these points corresponds a monomial in the hypersurface equation to which we associate a parameter $c_{i} \in \mathbb{C}$. Using SageMath we can finally recast this equation into the Weierstrass form
	\begin{equation}
	\label{eqWei}
		y^{2} = x^{3} + f(s,t)xz^{4} + g(s,t) z^{6}
	\end{equation}
where the homogeneous coordinates of the fiber are now $[x,y,z]$ in $\mathbb{P}^{(2,3,1)}$, $f$ and $g$ are respectively polynomials of degree 8 and 12 in $(s,t)$. The discriminant of \eqref{eqWei} is then $\mathit{\Delta}_{(f,g)} = 4f^{3} + 27 g^{2}$ and vanishes at 24 points which are the locations of 7-branes.

Once one has the Weierstrass form of the elliptically fibered K3, one finds the ADE groups associated to the various singularities using Kodaira and Neron classification \cite{kodaira_compact_1963,neron_modeles_1964}. Moreover, the moduli can be expressed via the parameters $c_{i}$ as shown in section \ref{sectionInvariant}. Sending those parameters to zero, we can therefore find possible enhancements of the group associated to a generic point in the moduli space. The third SageMath program then gives all possible enhancements obtained by sending all possible combinations of parameters $c_{i}$, when the hypersurface still defines a elliptically fibered K3.
\subsection{Graphs of Polytopes}
\label{graph}
Using this we construct \textit{graphs} of K3 surfaces, generalising the "chains" defined by Kreuzer and Skarke in \cite{kreuzer_classification_1998}. Nodes on a graph correspond to polytopes, or equivalently their associated K3 surface. We then link two polytopes if, by sending the same coefficient $c_{i}$ of \eqref{hypersurfaceEquation2} in every hypersurface equations for every possible fibration, we obtain the Weierstrass models of fibers of the other polytope\footnote{One does not necessarily obtain all the fibers of the polytope with fewer moduli. This is however the case for Figure \ref{enhancement}.}. Some of these graphs are represented in Figure \ref{pyramid1}, \ref{pyramid2} and \ref{pyramid3} and are discussed below. A less trivial case will be discussed in section \ref{section4}.

Let us consider Figure \ref{pyramid1}: $M0$ is linked to both $M5$ and $M6$ by which we mean that if one removes a particular point in the polytopes $M5$ and $M6$, one recovers the Weierstrass models corresponding to fibers of $M0$. This means that the moduli spaces of elliptically fibered K3s corresponding to the fibrations of the polytopes $M5$ and $M6$ contains the moduli spaces of fibers of the polytope $M0$. 

Figures \ref{pyramid1}, \ref{pyramid2} and \ref{pyramid3}, combined with the polytopes $M15,M30,M38,M104,M117$ with Picard 17 which are not linked to any polytope with higher Picard number, i.e. polytope with a lower number of moduli, describe all reflexive polyhedras up to 3 complex parameters.
 
\tikzstyle{debutfin}=[rounded corners,draw]

\begin{figure}[H]
\begin{center}
\begin{tikzpicture}[scale=1,every text node part/.style={align=center}]

\node[] (Picard18) at (-7,0) {\underline{Picard 19:}};
\node[] (Picard17) at (-7,-1.5) {\underline{Picard 18:}};
\node[] (Picard17) at (-7,-3) {\underline{Picard 17:}};

\node[draw,debutfin] (GP) at (0,0) {$M0$};
\node[draw,debutfin] (P1) at (-3,-1.5) {$M5$}; 
\node[draw,debutfin] (P2) at (3,-1.5) {$M6$};
\node[draw,debutfin] (C1) at (-5,-3) {$M21$};
\node[draw,debutfin] (C2) at (-3,-3) {$M26$};
\node[draw,debutfin] (C44) at (-1,-3) {$M28$};
\node[draw,debutfin] (C4) at (1,-3) {$M22$};
\node[draw,debutfin] (C5) at (3,-3) {$M24$};
\node[draw,debutfin] (C6) at (5,-3) {$M25$};

\draw[-] (GP) -- (P1);
\draw[-] (GP) -- (P2);

\draw[-] (P1) -- (C1);
\draw[-] (P1) -- (C2);
\draw[-] (P1) -- (C44);

\draw[-] (P2) -- (C4);
\draw[-] (P2) -- (C5);
\draw[-] (P2) -- (C6);
\draw[-] (P2) -- (C44);
\end{tikzpicture}
\caption{Polytopes up to 3 complex parameters that are linked to $M0$ by removing points in their $M$ lattice (i.e. a monomial in the hypersurface equation).}
\label{pyramid1}
\end{center}
\end{figure}
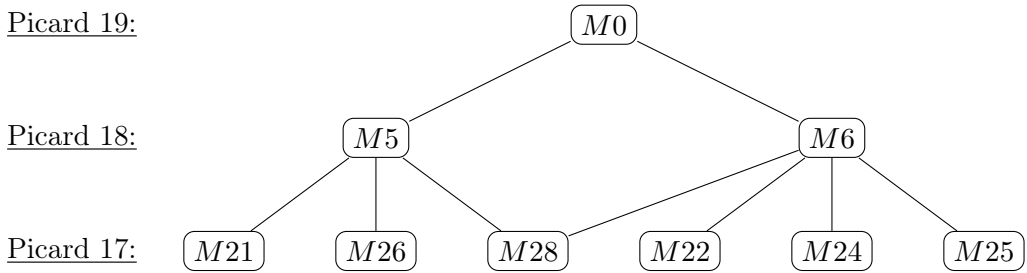

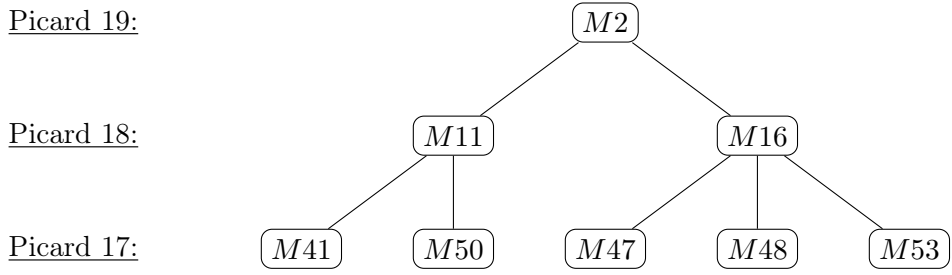
\begin{figure}[H]
\begin{center}
\begin{tikzpicture}[scale=1,every text node part/.style={align=center}]

\node[] (Picard18) at (-7,0) {\underline{Picard 19:}};
\node[] (Picard17) at (-7,-1.5) {\underline{Picard 18:}};
\node[] (Picard17) at (-7,-3) {\underline{Picard 17:}};
\node[] (P) at (5,0) {};

\node[draw,debutfin] (GP) at (0,0) {$M2$};
\node[draw,debutfin] (P1) at (-2,-1.5) {$M11$}; 
\node[draw,debutfin] (P2) at (2,-1.5) {$M16$};
\node[draw,debutfin] (C1) at (-4,-3) {$M41$};
\node[draw,debutfin] (C2) at (-2,-3) {$M50$};
\node[draw,debutfin] (C3) at (0, -3) {$M47$};
\node[draw,debutfin] (C5) at (2,-3) {$M48$};
\node[draw,debutfin] (C6) at (4,-3) {$M53$};

\draw[-] (GP) -- (P1);
\draw[-] (GP) -- (P2);

\draw[-] (P1) -- (C1);
\draw[-] (P1) -- (C2);

\draw[-] (P2) -- (C3);
\draw[-] (P2) -- (C5);
\draw[-] (P2) -- (C6);
\end{tikzpicture}
\end{center}
\caption{Polytopes up to 3 complex parameters that are linked to $M2$ by removing points in their $M$ lattice (i.e. a monomial in the hypersurface equation).}
\label{pyramid2}
\end{figure}

\begin{figure}[H]
\begin{center}
\begin{tikzpicture}[scale=1,every text node part/.style={align=center}]

\node[] (Picard18) at (-7,0) {\underline{Picard 18:}};
\node[] (Picard17) at (-7,-1.5) {\underline{Picard 17:}};

\node[draw,debutfin] (GP1) at (-5,0) {$M3$};
\node[draw,debutfin] (GP2) at (-2,0) {$M4$};
\node[draw,debutfin] (GP3) at (1,0) {$M7$};
\node[draw,debutfin] (GP4) at (3,0) {$M10$};
\node[draw,debutfin] (GP5) at (6,0) {$M88$};

\node[draw,debutfin] (P1) at (-5,-1.5) {$M27$};
\node[draw,debutfin] (P2) at (-3,-1.5) {$M20$};
\node[draw,debutfin] (P3) at (-1,-1.5) {$M29$};
\node[draw,debutfin] (P4) at (1,-1.5) {$M14$};
\node[draw,debutfin] (P5) at (3,-1.5) {$M49$};
\node[draw,debutfin] (P6) at (5,-1.5) {$M221$};
\node[draw,debutfin] (P7) at (7,-1.5) {$M230$};

\draw[-] (GP1) -- (P1) ;
\draw[-] (GP2) -- (P2) ;
\draw[-] (GP2) -- (P3) ;
\draw[-] (GP3) -- (P4) ;
\draw[-] (GP4) -- (P5) ;
\draw[-] (GP5) -- (P6) ;
\draw[-] (GP5) -- (P7) ;

\end{tikzpicture}
\end{center}
\caption{Links between polytopes with Picard 18 and 17. Going from Picard 17 to 18 amounts to removing a point the in the polytope in the $M$ lattice (i.e. a monomial in the hypersurface equation which defines the K3 surface).}
\label{pyramid3}
\end{figure}
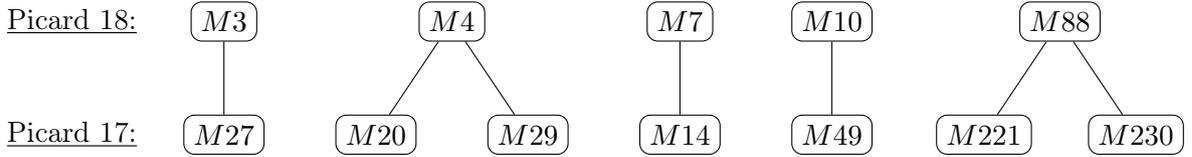

	\section{F-theory/Heterotic string duality in 8 dimensions: Wilson lines from Reflexive Polyhedras?}	
	\label{section4}	

	\subsection{Basic Aspects of Heterotic String}
	\label{heterotic}
	Here we introduce basic notions concerning the compactifications of heterotic strings on a two torus, based on \cite{blumenhagen_basic_2013,fraiman_new_2018}.
	As was presented in the beginning of the introduction, the moduli space of the heterotic string compactified on a two torus is given by the Narain space \eqref{narain}. This is due to the fact that the twenty dimensional internal momentum
	\begin{equation}
	\label{momentum}
	\textbf{P} = \left(\mathbf{p_{R}}, \mathbf{p_{L}}\right)  = \left(p_{R a}, \left(p_{L a}, p_{L}^{A}\right)\right)
	\end{equation}
with $a=1,2$ an index corresponding to directions along the two torus and $A=1,...,16$ an index along the 16 dimensional torus of heterotic string transforms as a vector under $O(2,18, \mathbb{R})$. One then obtains that $\mathbf{P}\cdot\mathbf{P} \in 2\mathbb{Z}$ and $\mathbf{P}$ forms a lorentzian lattice of signature $(2,18)$. The momentum \eqref{momentum} can be written in terms of the metric $G$ of the two torus, the two form field $B$, $U(1)^{16}$ gauge field $A_{a}^{A}$ as well as winding $w_{a}$ and momentum number $n_{a}$. When the Wilson lines $A_{a}^{A}$  are null, one obtains that an admissible state verifies $|p_{L}^{A}|^{2} \in 2\mathbb{Z}$, meaning that $p_{L}^{A}$ is lying in a 16 dimensional even self dual lattice\footnote{Self duality is due to modular invariance.}, i.e. correspond to the root vectors of $E_{8}\times E_{8}$ or the weight of $\frac{Spin(32)}{\mathbb{Z}_{2}}$. On the other hand, considering generic values for the two Wilson lines $A_{1}^{A}$ and $A_{2}^{A}$ completely breaks the gauge symmetry from $E_{8}\times E_{8}$ or $\frac{Spin(32)}{\mathbb{Z}_{2}}$ to $U(1)^{16}$ due to the quantization condition of the momentum number.

In \cite{fraiman_new_2018} is presented the compactifications of heterotic strings on a circle, with one Wilson line parametrized by one parameter. This leads to a wide variety of possible gauge groups one can obtain from heterotic strings. As an example one can consider one Wilson line of the form $A = (a_{8},0_{8})$ with $a\in \mathbb{R}$ breaking $SO(32)$ to $SO(16)\times SO(16)$ and $E_{8}\times E_{8}$ to $E_{7}\times E_{8}$ due to the quantization condition of the momentum number. Here we will write the Wilson lines $A_{1}$ and $A_{2}$ in a complex form $A = A_{1}+iA_{2}$. We then seek for possible Wilson lines parametrizations which will correspond to the gauge groups we obtain from elliptic fibrations of K3 surfaces. 
	\subsection{Graphs of Polytopes: from Monomials to Wilson lines...}
The duality map between F-theory on K3 and heterotic strings on a two torus has been explicitly written for two polytopes having each two inequivalent fibrations. The gauge groups associated to these fibers amazingly are $E_{8}\times E_{8}$ and $SO(32)$ for the first polytope ($M88$ using our notation) and $E_{7}\times E_{8}$ and $SO(28)\times SU(2)$ ($M221$). Using \cite{fraiman_new_2018}, we can see that adding a Wilson line of the form $A = (a_{2}, 0_{14})$ with $a \in \mathbb{C}$, using the notation of the previous section \ref{heterotic}  breaks  $E_{8}\times E_{8}$ to $E_{7}\times E_{8}$, and $SO(32)$ to $SO(28)\times SU(2)$ for a generic value of $a$. On the heterotic side one might be able to interpret the polytope $M221$ as a compactification on a two torus, together with  one Wilson line of the form $(a_{2}, 0_{14})$. In fact considering this particular parametrization of Wilson line, the enhancements one finds on both heterotic strings and F-theory exactly match, as was presented by Anamaria Font at CERN \cite{font_comments_2019}  and studied with more details in an upcomming paper \cite{font_comments_nodate-1}.
	
Now we want to see if we can make similar interpretations by considering polytopes which admit only two fibrations. Following the construction we presented in section \ref{graph} we seek a graph of polytopes with two fibers in its dual lattice and which contains $M88$. As an example let us consider the polytope $M1328$ which has Picard number 14. Its moduli space is parametrized by 6 complex parameters. We write the hypersurfaces equations $P_{G} = 0$ of its two fibers below, where $G$ is the group associated to its ADE singularities
\begin{dmath}
\label{eq1M1328}
P_{E_{6}\times SO(10)} = -\underline{c_{0}x_{0}x_{1}x_{2}x_{3}st}+\underline{c_{1}x_{2}x_{3}^{2}s}+c_{2}x_{0}^{2}x_{3}s+\underline{c_{3}x_{0}^{3}x_{1}st}+c_{4}x_{0}x_{1}^{3}x_{2}^{2}s^{2}t^{2}+c_{5}x_{1}^{2}x_{2}^{2}x_{3}t^{3}+\underline{c_{6}x_{1}^{4}x_{2}^{3}s^{2}t^{3}}+c_{7}x_{0}x_{1}^{3}x_{2}^{2}t^{4}+\underline{c_{8}x_{1}^{4}x_{2}^{3}t^{5}}+\underline{c_{9}x_{1}^{4}x_{2}^{3}st^{4}},
\end{dmath}	
\begin{dmath}
\label{eq2M1328}
P_{SU(11)\times SU(2)} = -\underline{c_{0}x_{0}x_{1}x_{2}x_{3}st}+\underline{c_{1}x_{1}x_{2}^{2}t}+c_{2}x_{0}x_{1}x_{2}x_{3}s^{2}+\underline{c_{3}x_{0}^{2}x_{1}x_{3}^{2}s^{3}}+c_{4}x_{1}^{2}x_{3}^{3}st^{4}+c_{5}x_{0}^{3}x_{2}+\underline{c_{6}x_{1}^{2}x_{3}^{3}t^{5}}+c_{7}x_{0}^{4}x_{3}s+\underline{c_{8}x_{0}^{4}x_{3}t}+\underline{c_{9}x_{0}^{2}x_{1}x_{3}^{2}t^{3}}
\end{dmath}
with $x_{i}$ homogeneous coordinates of the fiber, and $(s,t)$ coordinates on the base. These hypersurfaces can then be recast into a Weierstrass form where $(s,t)$ correspond to coordinates on the base $\mathbb{P}_{1}$\footnote{We do not write the Weierstrass models due to the size of the parameters $f$, $g$ and $\Delta_{(f,g)}.$} . Considering the underlined monomials in the equations \eqref{eq1M1328} and \eqref{eq2M1328} gives the Weierstrass models associated to the polytope $M88$ and thus corresponds to the heterotic strings without Wilson lines
	\begin{align}
	\label{e8e8}
	\begin{split}
	f =& \left(-\frac{1}{48}\right) \cdot t^{4} \cdot s^{4} \cdot c_{0}^{4}\\
g =& \left(-\frac{1}{864}\right) \cdot t^{5} \cdot s^{5} \cdot (864 c_{1}^{3}
c_{3}^{2} c_{6} s^{2} -  c_{0}^{6} s t + 864 c_{1}^{3} c_{3}^{2} c_{9} s
t + 864 c_{1}^{3} c_{3}^{2} c_{8} t^{2})\\
\Delta_{(f,g)} =& \left(\frac{1}{16}\right) \cdot c_{3}^{2} \cdot c_{1}^{3} \cdot t^{10}
\cdot s^{10} \cdot (c_{6} s^{2} + c_{9} s t + c_{8} t^{2}) \cdot (432
c_{1}^{3} c_{3}^{2} c_{6} s^{2} -  c_{0}^{6} s t \\&+ 432 c_{1}^{3}
c_{3}^{2} c_{9} s t + 432 c_{1}^{3} c_{3}^{2} c_{8} t^{2})
\end{split}
	\end{align}	
	for $E_{8}\times E_{8}$ and 
		\begin{align}
		\label{SO(32)}
		\begin{split}
	f =& \left(-\frac{1}{48}\right) \cdot t^{2} \cdot (16 c_{1}^{2} c_{3}^{2}
s^{6} - 8 c_{0}^{2} c_{1} c_{3} s^{5} t + c_{0}^{4} s^{4} t^{2} + 32
c_{1}^{2} c_{3} c_{9} s^{3} t^{3} - 8 c_{0}^{2} c_{1} c_{9} s^{2} t^{4}
\\&- 48 c_{1}^{2} c_{6} c_{8} t^{6} + 16 c_{1}^{2} c_{9}^{2} t^{6})\\
g =& \left(-\frac{1}{864}\right) \cdot t^{3} \cdot (4 c_{1} c_{3} s^{3} - 
c_{0}^{2} s^{2} t + 4 c_{1} c_{9} t^{3}) \cdot (16 c_{1}^{2} c_{3}^{2}
s^{6} - 8 c_{0}^{2} c_{1} c_{3} s^{5} t + c_{0}^{4} s^{4} t^{2} \\&+ 32
c_{1}^{2} c_{3} c_{9} s^{3} t^{3} - 8 c_{0}^{2} c_{1} c_{9} s^{2} t^{4}
- 72 c_{1}^{2} c_{6} c_{8} t^{6} + 16 c_{1}^{2} c_{9}^{2} t^{6})\\
\Delta_{(f,g)} =& \left(-\frac{1}{16}\right) \cdot c_{8}^{2} \cdot c_{6}^{2} \cdot
c_{1}^{4} \cdot t^{18} \cdot (16 c_{1}^{2} c_{3}^{2} s^{6} - 8 c_{0}^{2}
c_{1} c_{3} s^{5} t + c_{0}^{4} s^{4} t^{2} + 32 c_{1}^{2} c_{3} c_{9}
s^{3} t^{3}\\& - 8 c_{0}^{2} c_{1} c_{9} s^{2} t^{4} - 64 c_{1}^{2} c_{6}
c_{8} t^{6} + 16 c_{1}^{2} c_{9}^{2} t^{6})
	\end{split}
		\end{align}	
	for $SO(32)$. We can then define two moduli $\xi$ and $\rho$ via the equation \ref{moduli} 
	\begin{equation}
		\xi = \frac{c_{8}c_{6}}{c_{9}^{2}} \quad , \quad \eta = \frac{c_{9}c_{3}^{2}c_{1}^{3}}{c_{0}^{6}}.
	\end{equation}
 They parametrize the two dimensional moduli space and are found by considering linear relations on the edges of the polytope $M1328$ (see Equation \eqref{moduli})\footnote{They correspond to the parameter $u$ and $v$ in  \cite{candelas_type_2015}.}. 
  Now we know that adding a monomial corresponds to adding complex parameters in the Wilson lines. As we have to add four complex parameters which correspond to the four additional monomials $c_{2}$, $c_{4}$, $c_{5}$ and $c_{7}$ in \eqref{eq1M1328} and \eqref{eq2M1328}, we use the full graph which links $M1328$ to $M88$ represented in Figure \ref{enhancement}\footnote{This graph was found using the third program presented in this paper applied to the polytope $M1328$.}. Going down in the graph, we define four additional complex parameters as 
\begin{equation}
		\label{wilsonLine2}
		A_{c_{7}} = \frac{c_{7}c_{0}^{2}}{c_{1}c_{3}c_{8}} \quad , \quad
		A_{c_{2}} = \frac{c_{2}c_{1}}{c_{3}}  \quad , \quad 
		A_{c_{4}} = \frac{c_{4}c_{0}^{2}}{c_{1}c_{3}c_{6}} \quad , \quad
		A_{c_{5}} = \frac{c_{5}c_{0}^{3}}{c_{3}c_{8}c_{1}^{2}}.
\end{equation} 
We already know that the polytope $M221$ is obtained on the heterotic side by adding a Wilson line $a(1_{2},0_{14})$ therefore the monomial "$c_{7}$" is associated to this Wilson line. Looking at all the gauge groups in the graph and using results on the compactification of heterotic strings on a circle \cite{fraiman_new_2018} we find that a possibility for the Wilson lines associated to each monomial is
 \begin{align}
		\label{wilsonLine2}
		A_{c_{7}}\sim a(1_{2},0_{14})  \quad A_{c_{2}}  \sim b(1_{16})   \quad	A_{c_{4}} \sim c(0_{14},1_{2}) \quad 	A_{c_{5}} \sim   d\left[(1_{2},0_{14}) + i(0,1_{2},0_{13})\right]
	\end{align}
	with $a,b,c$ and $d$ in $\mathbb{C}$ parametrizing the moduli on the heterotic side. We can see that $A_{c_{7}}$ and $A_{c_{4}}$  are linked to the same Wilson line content if it were not for the symmetry breaking of $A_{c_{5}}$\footnote{In the $E_{8}\times E_{8}$ heterotic string one can just interchange the $E_{8}$s.}. Indeed if one does not add the monomial $c_{5}$, or the Wilson line $A_{c_{5}}$ in the dual theory, one can interchange $c_{7}$ and $c_{4}$ and obtain the same Weierstrass models obtained from $M221$. Moreover, due to the symmetry of the two parameters $A_{c_{7}}$ and $A_{c_{4}}$, if $A_{c_{7}} = A_{c_{4}}$ i.e $a=c$ in \eqref{wilsonLine2}, we obtain what we expect on the heterotic side, namely $SO(24)\times SU(2)^{2} \rightarrow SO(24)\times SU(4)$ for the polytope $M476$ while $E_{7}\times E_{7}$ is not enhanced. 
	
	\tikzset{myptr/.style={decoration={markings,mark=at position 1 with %
    {\arrow[scale=2,>=stealth]{>}}},postaction={decorate}}}
\tikzstyle{debutfin}=[rounded corners,draw]

	\begin{figure}
	\begin{center}
	\begin{tikzpicture}[scale=1.2,every text node part/.style={align=center}]
\node[draw,debutfin] (GP) at (0,0) {$M88$: $\binom{E_{8}\times E_{8}}{SO(32)}$};
\node[draw,debutfin] (P1) at (-2,-2) {$M221$: $\binom{E_{7}\times E_{8}}{SO(28)\times SU(2)}$}; 
\node[draw,debutfin] (P2) at (2,-2) {$M230$: $\binom{E_{7}\times E_{7}}{SU(16)}$};
\node[draw,debutfin] (C1) at (-4,-4) {$M473$: $\binom{E_{8}\times E_{6}}{SO(26)}$};
\node[draw,debutfin] (C2) at (0, -4) {$M497$: $\binom{E_{7}\times E_{6}}{SU(14)\times SU(2)}$};
\node[draw,debutfin] (C3) at (4,-4) {$M476$: $\binom{E_{7}\times E_{7}}{SO(24)\times SU(2)^{2}}$};
\node[draw,debutfin] (GC1) at (-4,-6) {$M859$: $\binom{E_{7}\times SO(10)}{SU(13)}$};
\node[draw,debutfin] (GC2) at (0, -6) {$M866$: $\binom{E_{7}\times E_{6}}{SO(22)\times SU(2)}$};
\node[draw,debutfin] (GC3) at (4,-6) {$M895$: $\binom{E_{6}\times E_{6}}{SU(12)\times SU(2)^{2}}$};
\node[draw,debutfin] (GGC) at (0,-8) {$M1328$: $\binom{E_{6}\times SO(10)}{SU(11)\times SU(2)}$};

\draw[myptr] (GP) -- (P1) node[midway,fill=white,text=red]{$A_{c_{7}} \neq 0$};
\draw[myptr] (GP) -- (P2) node[midway,fill=white,text=red]{$A_{c_{2}} \neq 0$};

\draw[myptr] (P1) -- (C1) node[midway,fill=white,text=red]{$A_{c_{5}}$};
\draw[myptr] (P1) -- (C2) node[midway,fill=white,text=red]{$A_{c_{2}}$};
\draw[myptr] (P1) -- (C3) node[near end,fill=white,text=red]{$A_{c_{4}}$};

\draw[myptr] (P2) -- (C2) node[near start,fill=white,text=red]{$A_{c_{7}}$};

\draw[myptr] (C1) -- (GC1) node[midway,fill=white,text=red] {$A_{c_{2}}$};
\draw[myptr] (C1) -- (GC2) node[near start,fill=white,text=red] {$A_{c_{4}}$};

\draw[myptr] (C2) -- (GC1) node[near start,fill=white,text=red] {$A_{c_{5}}$};
\draw[myptr] (C2) -- (GC3) node[near start,fill=white,text=red] {$A_{c_{4}}$};

\draw[myptr] (C3) -- (GC2) node[near start,fill=white,text=red] {$A_{c_{5}}$};
\draw[myptr] (C3) -- (GC3) node[midway,fill=white,text=red] {$A_{c_{2}}$};

\draw[myptr] (GC1) -- (GGC) node[midway,fill=white,text=red] {$A_{c_{4}}$};
\draw[myptr] (GC2) -- (GGC) node[midway,fill=white,text=red] {$A_{c_{2}}$};
\draw[myptr] (GC3) -- (GGC) node[midway,fill=white,text=red] {$A_{c_{5}}$};
	\end{tikzpicture}
	\end{center}
	\caption{Links between various reflexive polyhedras. Going upward from $M1328$ amounts to removing points in the polytope $M1328$, or equivalently monomials in the hypersurface equations \eqref{eq1M1328} and \eqref{eq2M1328}. Going downward corresponds to adding a complex modulus $A_{c_{\#}}$.}
	\label{enhancement}
	\end{figure}
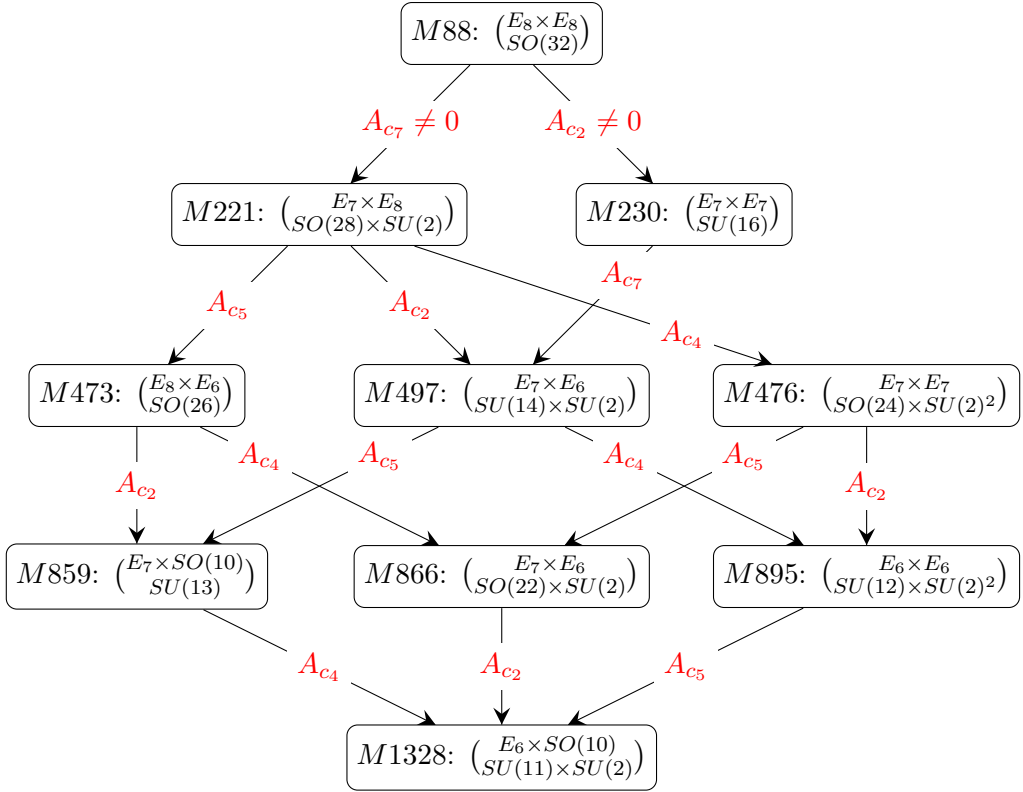

	\subsection{... and back to Monomials}
	
	We are now able for particular polytopes $M\#$ whose dual $N\#$ contain two fibers to describe K3s as parametrizations of Wilson lines of its dual theory (both for $E_{8}\times E_{8}$ and $SO(32)$). Rather we linked \textit{monomials} in the defining hypersurface equation of K3s to parameters in the Wilson lines. This means that we can construct Weierstrass models of elliptically fibered K3s which are not per say described by reflexive polyhedras, and directly interpret them as a particular Wilson lines content on the $E_{8}\times E_{8}$ and $SO(32)$ heterotic strings. Indeed let us go back to the graph of Figure \ref{enhancement}: adding the monomial $c_{4}$ to the underlined terms of \eqref{eq1M1328} and \eqref{eq2M1328} gives the Weierstrass models one gets from $M221$ as explained previously. Adding only $c_{5}$ however, we cannot obtain a polytope with 3 moduli which will give the same Weierstrass models. Thus let us write the Weierstrass models of the polytope $M88$, together with the additional monomial $c_{5}$ in \eqref{eq1M1328} and \eqref{eq2M1328}. For the first fiber we find
	\begin{dmath}
	f = \left(-\frac{1}{48}\right) \cdot c_{0} \cdot s^{3} \cdot t^{4} \cdot
(c_{0}^{3} s - 24 c_{1} c_{3} c_{5} t)
	\end{dmath}
	\begin{dmath}
	g = \left(-\frac{1}{864}\right) \cdot s^{4} \cdot t^{5} \cdot (864 c_{1}^{3}
c_{3}^{2} c_{6} s^{3} -  c_{0}^{6} s^{2} t + 864 c_{1}^{3} c_{3}^{2}
c_{9} s^{2} t + 36 c_{0}^{3} c_{1} c_{3} c_{5} s t^{2} + 864 c_{1}^{3}
c_{3}^{2} c_{8} s t^{2} - 216 c_{1}^{2} c_{3}^{2} c_{5}^{2} t^{3})
	\end{dmath}
	\begin{dmath}
\Delta_{(f,g)} = \left(\frac{1}{16}\right) \cdot c_{3}^{2} \cdot c_{1}^{3} \cdot s^{8}
\cdot t^{10} \cdot (432 c_{1}^{3} c_{3}^{2} c_{6}^{2} s^{6} -  c_{0}^{6}
c_{6} s^{5} t + 864 c_{1}^{3} c_{3}^{2} c_{6} c_{9} s^{5} t + 36
c_{0}^{3} c_{1} c_{3} c_{5} c_{6} s^{4} t^{2} + 864 c_{1}^{3} c_{3}^{2}
c_{6} c_{8} s^{4} t^{2} -  c_{0}^{6} c_{9} s^{4} t^{2} + 432 c_{1}^{3}
c_{3}^{2} c_{9}^{2} s^{4} t^{2} - 216 c_{1}^{2} c_{3}^{2} c_{5}^{2}
c_{6} s^{3} t^{3} -  c_{0}^{6} c_{8} s^{3} t^{3} + 36 c_{0}^{3} c_{1}
c_{3} c_{5} c_{9} s^{3} t^{3} + 864 c_{1}^{3} c_{3}^{2} c_{8} c_{9}
s^{3} t^{3} + 36 c_{0}^{3} c_{1} c_{3} c_{5} c_{8} s^{2} t^{4} + 432
c_{1}^{3} c_{3}^{2} c_{8}^{2} s^{2} t^{4} - 216 c_{1}^{2} c_{3}^{2}
c_{5}^{2} c_{9} s^{2} t^{4} -  c_{0}^{3} c_{3} c_{5}^{3} s t^{5} - 216
c_{1}^{2} c_{3}^{2} c_{5}^{2} c_{8} s t^{5} + 27 c_{1} c_{3}^{2}
c_{5}^{4} t^{6})
\end{dmath}
and for the second
	\begin{dmath}
f = \left(-\frac{1}{48}\right) \cdot t^{2} \cdot (16 c_{1}^{2} c_{3}^{2}
s^{6} - 8 c_{0}^{2} c_{1} c_{3} s^{5} t + c_{0}^{4} s^{4} t^{2} + 32
c_{1}^{2} c_{3} c_{9} s^{3} t^{3} - 8 c_{0}^{2} c_{1} c_{9} s^{2} t^{4}
- 24 c_{0} c_{1} c_{5} c_{6} s t^{5} - 48 c_{1}^{2} c_{6} c_{8} t^{6} +
16 c_{1}^{2} c_{9}^{2} t^{6})
\end{dmath}
\begin{dmath}
g = \left(-\frac{1}{864}\right) \cdot t^{3} \cdot (64 c_{1}^{3} c_{3}^{3}
s^{9} - 48 c_{0}^{2} c_{1}^{2} c_{3}^{2} s^{8} t + 12 c_{0}^{4} c_{1}
c_{3} s^{7} t^{2} -  c_{0}^{6} s^{6} t^{3} + 192 c_{1}^{3} c_{3}^{2}
c_{9} s^{6} t^{3} - 96 c_{0}^{2} c_{1}^{2} c_{3} c_{9} s^{5} t^{4} - 144
c_{0} c_{1}^{2} c_{3} c_{5} c_{6} s^{4} t^{5} + 12 c_{0}^{4} c_{1} c_{9}
s^{4} t^{5} + 36 c_{0}^{3} c_{1} c_{5} c_{6} s^{3} t^{6} - 288 c_{1}^{3}
c_{3} c_{6} c_{8} s^{3} t^{6} + 192 c_{1}^{3} c_{3} c_{9}^{2} s^{3}
t^{6} + 72 c_{0}^{2} c_{1}^{2} c_{6} c_{8} s^{2} t^{7} - 48 c_{0}^{2}
c_{1}^{2} c_{9}^{2} s^{2} t^{7} - 144 c_{0} c_{1}^{2} c_{5} c_{6} c_{9}
s t^{8} - 216 c_{1}^{2} c_{5}^{2} c_{6}^{2} t^{9} - 288 c_{1}^{3} c_{6}
c_{8} c_{9} t^{9} + 64 c_{1}^{3} c_{9}^{3} t^{9})
\end{dmath}
\begin{dmath}
\Delta_{(f,g)} = \left(-\frac{1}{16}\right) \cdot c_{6}^{2} \cdot c_{1}^{3} \cdot t^{15}
\cdot (16 c_{1}^{2} c_{3}^{3} c_{5}^{2} s^{9} - 8 c_{0}^{2} c_{1}
c_{3}^{2} c_{5}^{2} s^{8} t + c_{0}^{4} c_{3} c_{5}^{2} s^{7} t^{2} + 16
c_{0} c_{1}^{2} c_{3}^{2} c_{5} c_{8} s^{7} t^{2} - 8 c_{0}^{3} c_{1}
c_{3} c_{5} c_{8} s^{6} t^{3} + 16 c_{1}^{3} c_{3}^{2} c_{8}^{2} s^{6}
t^{3} + 48 c_{1}^{2} c_{3}^{2} c_{5}^{2} c_{9} s^{6} t^{3} + c_{0}^{5}
c_{5} c_{8} s^{5} t^{4} - 8 c_{0}^{2} c_{1}^{2} c_{3} c_{8}^{2} s^{5}
t^{4} - 16 c_{0}^{2} c_{1} c_{3} c_{5}^{2} c_{9} s^{5} t^{4} - 36 c_{0}
c_{1} c_{3} c_{5}^{3} c_{6} s^{4} t^{5} + c_{0}^{4} c_{1} c_{8}^{2}
s^{4} t^{5} + c_{0}^{4} c_{5}^{2} c_{9} s^{4} t^{5} + 32 c_{0} c_{1}^{2}
c_{3} c_{5} c_{8} c_{9} s^{4} t^{5} + c_{0}^{3} c_{5}^{3} c_{6} s^{3}
t^{6} - 72 c_{1}^{2} c_{3} c_{5}^{2} c_{6} c_{8} s^{3} t^{6} - 8
c_{0}^{3} c_{1} c_{5} c_{8} c_{9} s^{3} t^{6} + 32 c_{1}^{3} c_{3}
c_{8}^{2} c_{9} s^{3} t^{6} + 48 c_{1}^{2} c_{3} c_{5}^{2} c_{9}^{2}
s^{3} t^{6} - 30 c_{0}^{2} c_{1} c_{5}^{2} c_{6} c_{8} s^{2} t^{7} - 8
c_{0}^{2} c_{1}^{2} c_{8}^{2} c_{9} s^{2} t^{7} - 8 c_{0}^{2} c_{1}
c_{5}^{2} c_{9}^{2} s^{2} t^{7} - 96 c_{0} c_{1}^{2} c_{5} c_{6}
c_{8}^{2} s t^{8} - 36 c_{0} c_{1} c_{5}^{3} c_{6} c_{9} s t^{8} + 16
c_{0} c_{1}^{2} c_{5} c_{8} c_{9}^{2} s t^{8} - 27 c_{1} c_{5}^{4}
c_{6}^{2} t^{9} - 64 c_{1}^{3} c_{6} c_{8}^{3} t^{9} - 72 c_{1}^{2}
c_{5}^{2} c_{6} c_{8} c_{9} t^{9} + 16 c_{1}^{3} c_{8}^{2} c_{9}^{2}
t^{9} + 16 c_{1}^{2} c_{5}^{2} c_{9}^{3} t^{9})
\end{dmath}
	 The gauge groups associated to the singularities of these Weierstrass models are $E_{6}\times E_{8}$ and $SO(26)$ respectively. They are exactly what we expect from heterotic string theories with one Wilson line $A_{c_{5}}$ in equation \eqref{wilsonLine2}. This means that if we compactify F-theory on these elliptically fibered K3s, we know that the Wilson line content on the dual heterotic strings should be of a similar kind as $A_{c_{5}}$. Using this it is then possible to restrict the study of the duality map between the two theories to a three dimensional moduli space to verify that the enhancements on both F-theory and heterotic sides match. 	
\subsection{Wilson line Interpretation for Polytope with more than Two Fibers}
		The  Wilson line description of reflexive polyhedras can be extended to K3 surfaces which have more than two inequivalent elliptic fibrations. Indeed let us consider the polytope $M2$ with three fibers presented in the Figure \ref{pic19}. The fiber $E_{8}\times E_{8}\times SU(2)$ is obtained via the fiber $E_{8}\times E_{8}$ of the polytope $M88$ with $\xi = \frac{1}{4}$ \cite{candelas_type_2015}. This in fact corresponds to taking the complex structure and Kahler moduli equal when compactifying on the two torus on the heterotic string. The two remaining fibers ($E_{7}\times SO(20)$ and $SU(18)$) of $M2$ can be obtained by considering $M1328$ with $c_{3} = c_{4} = c_{7} = c_{8} = c_{9} = 0$: $E_{6}\times SO(10)$ is enhanced to $E_{7}\times SO(20)$ while $SU(11)\times SU(2)$ to $SU(18)$. From our construction, the Wilson lines on the dual theory are therefore parametrized by $A_{c_{5}}$ and $A_{c_{2}}$. The moduli spaces for the fibers $E_{7}\times SO(20)$ and $SU(18)$ are thus contained in the moduli spaces of the heterotic strings $E_{8}\times E_{8}$ and $SO(32)$ with this particular Wilson lines parametrization respectively. 
	\section{Conclusion}
	 Here we showed how to interpret particular elliptically fibered K3 surfaces directly from a Wilson line parametrization of the dual $E_{8}\times E_{8}$ and $SO(32)$ heterotic strings. We constructed a graph of polytopes with two fibers where the links can be considered in this particular case as inclusion relations between the moduli spaces of the elliptically fibered K3s associated to each fiber. Because in some limit of the moduli space we now recover the compactifications of F-theory dual to the ones of heterotic strings on a two torus with no Wilson lines, we can interpret the additional complex parameters in the moduli spaces as being dual to Wilson line moduli. They appear in the compactifications of F-theory as additional monomials in the hypersurface equation defining the K3 surface. Therefore there seem to be a close link between monomials on the F-theory side, and Wilson line moduli in the heterotic theories. Adding only one monomial to the hypersurface equations which correspond to the $E_{8}\times E_{8}$ and $SO(32)$ compactifications, we can restrict the number of complex parameters in the moduli space to three i.e. one Wilson line moduli for the heterotic strings. This makes finding possible enhancements more convenient, and a study similar to the ones presented in \cite{font_comments_2019} and upcoming paper \cite{font_comments_nodate-1} should give more insights on the Wilson line description of elliptically fibered K3 surfaces. Moreover, we also showed that this construction can be adapted to polytopes with more than two fibers. We were able to find heterotic duals to the three fibers of the polytope $M2$ with one moduli. Using the various SageMath programs we developed as well as graphs of polytopes, we hope to find more insights on K3s with three fibrations and more. Finally here we focused on compactifications of F-theory and heterotic strings to eight dimensions. We expect that the monomial and Wilson line moduli duality should be valuable in studying compactifications to lower dimensions, both involving K3 surfaces and more generally Calabi-Yau threefolds.
	 
\acknowledgments{I would like to thank my PhD supervisor Mariana Graña for her guidance on this project. I am grateful to Anamaria Font, Christoph Mayrhofer and Hector Parra for sharing their notes on their current work, and for helpful discussions concerning the constructions of Weierstrass models using Sagemath. I would also like to thank Bernado Fraiman and Carmen Nuñez for their clarification on heterotic compactifications. This work was supported in part by the Ecole Doctorale Physique en Île-de-France grant and the ERC Consolidator Grant 772408-Stringlandscape.}
\appendix
\section{Program 1: Dynkin Diagram from Reflexive Polyhedra}
\label{appendixPic}
Here we present how to use the first program. The first line is simply \textit{reflexivePolytopes = []}. Just enter a list of number between 0 and 4318 to consider the reflexive polytopes \textit{ReflexivePolytope(3,\#)} of this list into Sagemath. The program then returns a table containing all gauge groups for all the fibers of any reflexive polytope. The table is written in latex format on a text file.

Figure \ref{picard17} shows the output of this first program where we gave as an entry the reflexive polytope with Picard number 17 (i.e. 3 complex parameters) and no correction term.

\begin{landscape}
\begin{figure}
\caption{Picard 17}
\label{picard17}
\begin{adjustbox}{max width=\linewidth}
\begin{tabular}{|c|c|c|c|c|c|c|c|c|c|c|}
\hline
N14 & $\frac{E_{7} \times SU(8)\times SU(2)}{Z_{2}}$ &$SO(16) \times E_{6}$  &$E_{8} \times E_{7}\times Z_{4}$  &$SU(14) \times SU(2)$  &$E_{8} \times E_{6}$  &$SO(10) \times SO(18)$  &\\ \hline
N15 & $E_{7} \times E_{6}\times SU(2)$  &$SO(14) \times SO(14)$  &$\frac{SO(16) \times SU(8)}{Z_{2}}$  &$E_{7} \times E_{8}\times Z_{4}$  &$\frac{SU(14) \times SU(2)}{Z_{2}}$  &\\ \hline
N20 & $E_{7} \times E_{8}\times Z_{3}$  &$E_{6} \times SO(14)$  &$E_{6} \times SO(10)\times SU(3)SU(2)$  &$SU(6) \times SO(14)\times SU(3)$  &$E_{7} \times SO(12)$  &$SO(12) \times SU(8)$  &$\frac{SU(10) \times SO(8)\times SU(2)}{Z_{2}}$ &$SU(9) \times SU(7)$  &\\ \hline
N21 & $E_{7} \times SO(12)\times SU(2)$  &$SU(6) \times SU(10)$  &$E_{7} \times E_{7}$  &$\frac{SO(12) \times SO(12)\times SU(2)SU(2)}{Z_{2}}$ &$E_{7} \times E_{7}\times Z_{2}$  &$SU(8) \times E_{6}$  &$\frac{SO(12) \times SO(12)\times SU(4)}{Z_{2}}$ &$\frac{SO(16) \times SO(8)\times SU(2)SU(2)}{Z_{2}}$ &\\ \hline
N22 & $SO(10) \times SU(9)\times SU(2)$  &$SO(14) \times SU(7)$  &$E_{8} \times E_{7}\times Z_{3}$  &$SO(14) \times SO(12)$  &$E_{6} \times E_{6}\times SU(3)$  &$SU(10) \times SO(8)$  &$E_{7} \times SO(10)\times SU(2)$  &$\frac{E_{6} \times SU(6)\times SU(3)SU(3)}{Z_{3}}$ &\\ \hline
N23 & $\frac{SU(10) \times SO(12)}{Z_{2}}$  &$E_{6} \times SO(14)\times SU(2)$  &$E_{7} \times SU(8)$  &$SO(12) \times E_{8}$  &$E_{7} \times E_{7}$  &$SU(3) \times SU(13)$  &$SO(18) \times SU(6)$  &\\ \hline
N24 & $E_{7} \times E_{7}$  &$SO(10) \times SU(8)\times SU(3)$  &$E_{6} \times E_{6}\times SU(2)$  &$SO(12) \times SO(12)$  &$\frac{SU(8) \times SU(8)}{Z_{2}}$  &\\ \hline
N25 & $SO(10) \times SO(14)\times SU(2)$  &$SO(10) \times SU(9)$  &$E_{6} \times E_{7}$  &$SO(12) \times E_{6}\times SU(3)$  &$E_{7} \times E_{7}$  &$E_{8} \times E_{7}\times Z_{3}$  &$SU(7) \times E_{6}\times SU(3)$  &$\frac{SU(6) \times SU(9)\times SU(3)}{Z_{3}}$ &$\frac{SU(8) \times SO(12)\times SU(2)}{Z_{2}}$ &\\ \hline
N26 & $SU(9) \times E_{6}$  &$E_{6} \times SU(9)$  &$E_{7} \times E_{7}$  &$SO(10) \times SO(16)$  &$E_{6} \times E_{8}$  &$\frac{SU(6) \times E_{7}\times SU(4)}{Z_{2}}$ &$\frac{SU(3) \times SU(12)\times SU(3)}{Z_{3}}$ &\\ \hline
N27 & $SU(8) \times SO(14)$  &$\frac{SU(4) \times SU(12)}{Z_{2}}$  &$E_{6} \times E_{6}\times SU(2)SU(2)$  &$E_{7} \times SO(12)$  &\\ \hline
N28 & $E_{6} \times E_{7}\times SU(2)$  &$SU(10) \times SO(10)$  &$SO(10) \times E_{7}\times SU(3)$  &$SU(5) \times SU(11)$  &$SO(14) \times SO(12)$  &$E_{6} \times E_{8}$  &$E_{6} \times SU(8)\times SU(2)$  &$\frac{SU(6) \times SO(16)\times SU(2)}{Z_{2}}$ &\\ \hline
N29 & $E_{7} \times E_{6}$  &$E_{7} \times E_{7}\times Z_{2}$  &$\frac{SO(12) \times SU(8)\times SU(2)}{Z_{2}}$ &$\frac{SU(6) \times SU(10)}{Z_{2}}$  &$SO(12) \times E_{6}\times SU(2)$  &$SO(10) \times SO(14)\times SU(3)$  &$SU(8) \times SU(8)$  &\\ \hline
N30 & $E_{6} \times E_{6}$  &$E_{7} \times E_{7}\times Z_{2}$  &$\frac{SO(12) \times SO(12)\times SU(2)SU(2)}{Z_{2}}$ &$\frac{SU(8) \times SU(8)}{Z_{2}}$  &\\ \hline
N38 & $E_{7} \times E_{7}$  &$SO(14) \times E_{8}$  &$\frac{SU(4) \times SO(24)}{Z_{2}}$  &$\frac{U1 \times SU(16)}{Z_{2}}$  &\\ \hline
N41 & $\frac{SO(12) \times SO(16)\times SU(2)}{Z_{2}}$ &$E_{7} \times E_{8}\times Z_{4}$  &$\frac{SO(12) \times E_{7}\times SU(2)SU(2)}{Z_{2}}$ &$E_{7} \times E_{7}\times SU(2)$  &$U1 \times SU(14)$  &\\ \hline
N47 & $SO(18) \times SO(10)$  &$E_{8} \times E_{7}\times Z_{4}$  &$SU(14) \times U1\times SU(2)$  &$E_{8} \times E_{6}$  &$SO(16) \times E_{6}$  &$\frac{SU(8) \times E_{7}\times SU(2)}{Z_{2}}$ &\\ \hline
N48 & $E_{7} \times E_{7}$  &$SO(14) \times E_{6}\times SU(2)$  &$U1 \times SU(13)\times SU(3)$  &$\frac{SU(10) \times SO(12)}{Z_{2}}$  &\\ \hline
N49 & $E_{7} \times E_{6}\times SU(2)$  &$\frac{SU(8) \times SO(16)}{Z_{2}}$  &$E_{7} \times E_{8}\times Z_{4}$  &$SO(14) \times SO(14)$  &$\frac{SU(14) \times U1\times SU(2)}{Z_{2}}$ &\\ \hline
N50 & $SU(15) \times U1$  &$\frac{SU(6) \times SO(20)}{Z_{2}}$  &$SO(14) \times E_{7}$  &$E_{8} \times E_{6}\times SU(2)$  &\\ \hline
N53 & $E_{6} \times SU(9)$  &$E_{7} \times E_{7}$  &$SO(16) \times SO(10)$  &$\frac{U1 \times SU(12)\times SU(3)SU(3)}{Z_{3}}$ &\\ \hline
N104 & $E_{7} \times E_{8}$  &$\frac{SO(28) \times SU(2)}{Z_{2}}$  &\\ \hline
N117 & $\frac{SO(24) \times U1\times SU(4)}{Z_{2}}$ &$\frac{SU(16) \times U1}{Z_{2}}$  &$E_{7} \times E_{7}$  &\\ \hline
N221 & $E_{7} \times E_{8}$  &$\frac{SO(28) \times U1\times SU(2)}{Z_{2}}$ &\\ \hline
N230 & $\frac{SU(16) \times U1}{Z_{2}}$  &$E_{7} \times E_{7}$  &\\ \hline
\end{tabular}
\end{adjustbox}
\end{figure}
\end{landscape}

\section{Program 2: Weierstrass Models}
\label{program2}
Here we present again the typical output of the second computer program. Again on the first line one just specifies in a list the reflexive polytopes \# (associated to \textit{ReflexivePolytope(3,\#)}. The output is the hypersurface equation for every fibration of the K3 surface as well as the corresponding Weierstrass models (upon a choice fiber described in Figure \ref{polytopes} for F13, F15 and F16. In another file are saved all the hypersurface equations in Sagemath form.\\
The following is the typical Latex output when putting as an input "[476]".\\
\begin{center}
\Large \textbf{\underline{Polytope M476}}
\end{center}
\normalsize
Number of different Fiber is 2\\ 
\begin{center} 
 \large Fiber 1\normalsize 
 \end{center}  
\underline{The hypersurface equation is:} 
\begin{dmath}
p = -c_{0}x_{0}x_{1}x_{2}st+c_{1}x_{0}^{2}+c_{2}x_{2}^{3}+c_{3}x_{1}^{4}x_{2}s^{3}t^{5}+c_{4}x_{1}^{4}x_{2}s^{5}t^{3}+c_{5}x_{1}^{6}s^{7}t^{5}+c_{6}x_{1}^{6}s^{5}t^{7}+c_{7}x_{1}^{6}s^{6}t^{6}+
\end{dmath}
\underline{Data of the Weierstrass model:} 
\begin{dmath}
f = \left(\frac{1}{48}\right) \cdot t^{3} \cdot s^{3} \cdot (48 c_{1}^{2} c_{2} c_{4} s^{2} -  c_{0}^{4} s t + 48 c_{1}^{2} c_{2} c_{3} t^{2}) 
\end{dmath}
\begin{dmath}
g = \left(-\frac{1}{864}\right) \cdot t^{5} \cdot s^{5} \cdot (72 c_{0}^{2} c_{1}^{2} c_{2} c_{4} s^{2} + 864 c_{1}^{3} c_{2}^{2} c_{5} s^{2} -  c_{0}^{6} s t + 864 c_{1}^{3} c_{2}^{2} c_{7} s t + 72 c_{0}^{2} c_{1}^{2} c_{2} c_{3} t^{2} + 864 c_{1}^{3} c_{2}^{2} c_{6} t^{2}) 
\end{dmath}
\begin{dmath}
\Delta_{(f,g)} = \left(\frac{1}{16}\right) \cdot c_{2}^{2} \cdot c_{1}^{3} \cdot t^{9} \cdot s^{9} \cdot (64 c_{1}^{3} c_{2} c_{4}^{3} s^{6} -  c_{0}^{4} c_{1} c_{4}^{2} s^{5} t + 72 c_{0}^{2} c_{1}^{2} c_{2} c_{4} c_{5} s^{5} t + 432 c_{1}^{3} c_{2}^{2} c_{5}^{2} s^{5} t + 192 c_{1}^{3} c_{2} c_{3} c_{4}^{2} s^{4} t^{2} -  c_{0}^{6} c_{5} s^{4} t^{2} + 72 c_{0}^{2} c_{1}^{2} c_{2} c_{4} c_{7} s^{4} t^{2} + 864 c_{1}^{3} c_{2}^{2} c_{5} c_{7} s^{4} t^{2} - 2 c_{0}^{4} c_{1} c_{3} c_{4} s^{3} t^{3} + 72 c_{0}^{2} c_{1}^{2} c_{2} c_{3} c_{5} s^{3} t^{3} + 72 c_{0}^{2} c_{1}^{2} c_{2} c_{4} c_{6} s^{3} t^{3} + 864 c_{1}^{3} c_{2}^{2} c_{5} c_{6} s^{3} t^{3} -  c_{0}^{6} c_{7} s^{3} t^{3} + 432 c_{1}^{3} c_{2}^{2} c_{7}^{2} s^{3} t^{3} + 192 c_{1}^{3} c_{2} c_{3}^{2} c_{4} s^{2} t^{4} -  c_{0}^{6} c_{6} s^{2} t^{4} + 72 c_{0}^{2} c_{1}^{2} c_{2} c_{3} c_{7} s^{2} t^{4} + 864 c_{1}^{3} c_{2}^{2} c_{6} c_{7} s^{2} t^{4} -  c_{0}^{4} c_{1} c_{3}^{2} s t^{5} + 72 c_{0}^{2} c_{1}^{2} c_{2} c_{3} c_{6} s t^{5} + 432 c_{1}^{3} c_{2}^{2} c_{6}^{2} s t^{5} + 64 c_{1}^{3} c_{2} c_{3}^{3} t^{6}) 
\end{dmath}
\hrule 
\begin{center} 
 \large Fiber 2\normalsize 
 \end{center}  
\underline{The hypersurface equation is:} 
\begin{dmath}
p = -c_{0}x_{0}x_{1}x_{2}st+c_{1}x_{0}^{2}+c_{2}x_{1}^{2}x_{2}^{2}st^{3}+c_{3}x_{1}^{4}x_{2}t+c_{4}x_{2}^{3}s^{6}t+c_{5}x_{2}^{3}s^{7}+c_{6}x_{1}^{4}x_{2}s+c_{7}x_{1}^{2}x_{2}^{2}s^{4}+
\end{dmath}
\underline{Data of the Weierstrass model:} 
\begin{dmath}
f = \left(-\frac{1}{48}\right) \cdot s^{2} \cdot (-48 c_{1}^{2} c_{5} c_{6} s^{6} + 16 c_{1}^{2} c_{7}^{2} s^{6} - 48 c_{1}^{2} c_{3} c_{5} s^{5} t - 48 c_{1}^{2} c_{4} c_{6} s^{5} t - 48 c_{1}^{2} c_{3} c_{4} s^{4} t^{2} - 8 c_{0}^{2} c_{1} c_{7} s^{4} t^{2} + 32 c_{1}^{2} c_{2} c_{7} s^{3} t^{3} + c_{0}^{4} s^{2} t^{4} - 8 c_{0}^{2} c_{1} c_{2} s t^{5} + 16 c_{1}^{2} c_{2}^{2} t^{6}) 
\end{dmath}
\begin{dmath}
g = \left(-\frac{1}{864}\right) \cdot s^{3} \cdot (4 c_{1} c_{7} s^{3} -  c_{0}^{2} s t^{2} + 4 c_{1} c_{2} t^{3}) \cdot (-72 c_{1}^{2} c_{5} c_{6} s^{6} + 16 c_{1}^{2} c_{7}^{2} s^{6} - 72 c_{1}^{2} c_{3} c_{5} s^{5} t - 72 c_{1}^{2} c_{4} c_{6} s^{5} t - 72 c_{1}^{2} c_{3} c_{4} s^{4} t^{2} - 8 c_{0}^{2} c_{1} c_{7} s^{4} t^{2} + 32 c_{1}^{2} c_{2} c_{7} s^{3} t^{3} + c_{0}^{4} s^{2} t^{4} - 8 c_{0}^{2} c_{1} c_{2} s t^{5} + 16 c_{1}^{2} c_{2}^{2} t^{6}) 
\end{dmath}
\begin{dmath}
\Delta_{(f,g)} = \left(-\frac{1}{16}\right) \cdot c_{1}^{4} \cdot s^{14} \cdot (c_{6} s + c_{3} t)^{2} \cdot (c_{5} s + c_{4} t)^{2} \cdot (-64 c_{1}^{2} c_{5} c_{6} s^{6} + 16 c_{1}^{2} c_{7}^{2} s^{6} - 64 c_{1}^{2} c_{3} c_{5} s^{5} t - 64 c_{1}^{2} c_{4} c_{6} s^{5} t - 64 c_{1}^{2} c_{3} c_{4} s^{4} t^{2} - 8 c_{0}^{2} c_{1} c_{7} s^{4} t^{2} + 32 c_{1}^{2} c_{2} c_{7} s^{3} t^{3} + c_{0}^{4} s^{2} t^{4} - 8 c_{0}^{2} c_{1} c_{2} s t^{5} + 16 c_{1}^{2} c_{2}^{2} t^{6}) 
\end{dmath}
\hrule 
\normalsize
\section{Program 3: Finding Basic Enhancements and Constructing Graphs}
\label{program3}
In the following we see the enhancement for the input [476] for the third program.\\
\begin{center}
\Large $\mathbf{M476}$ \normalsize
\end{center}
\begin{multicols}{2}
\textbf{fiber 1}\\
()- E7xE7\\ 
(0_)- E7xE7\\ 
(3_)- E8xE7\\ 
(4_)- E7xE8\\ 
(5_)- E7xE7\\ 
(6_)- E7xE7\\ 
(7_)- E7xE7\\ 
(0_3_)- E8xE7\\ 
(0_4_)- E7xE8\\ 
(0_5_)- E7xE7\\ 
(0_6_)- E7xE7\\ 
(0_7_)- E7xE7\\ 
(3_4_)- E8xE8\\ 
(3_5_)- E8xE7\\ 
(3_7_)- E8xE7\\ 
(4_6_)- E7xE8\\ 
(4_7_)- E7xE8\\ 
(5_6_)- E7xE7\\ 
(5_7_)- E7xE7\\ 
(6_7_)- E7xE7\\ 
(0_3_4_)- E8xE8\\ 
(0_3_5_)- E8xE7\\ 
(0_3_7_)- E8xE7\\ 
(0_4_6_)- E7xE8\\ 
(0_4_7_)- E7xE8\\ 
(0_5_6_)- E7xE7\\ 
(0_5_7_)- E7xE7\\ 
(0_6_7_)- E7xE7\\ 
(3_4_7_)- E8xE8\\ 
(3_5_7_)- E8xE7\\ 
(4_6_7_)- E7xE8\\ 
(5_6_7_)- E7xE7\\ 
(0_3_4_7_)- E8xE8\\ 
(0_3_5_7_)- E8xE7\\ 
(0_4_6_7_)- E7xE8\\ 
(0_5_6_7_)- E7xE7\\  \\
\textbf{fiber 2}\\
()- SO(24)xSU(2)xSU(2)\\ 
(0_)- SO(24)xSU(2)xSU(2)\\ 
(3_)- SO(28)xSU(2)\\ 
(4_)- SO(28)xSU(2)\\ 
(5_)- SO(24)xSU(2)xSU(2)\\ 
(6_)- SO(24)xSU(2)xSU(2)\\ 
(7_)- SO(24)xSU(2)xSU(2)\\ 
(0_3_)- SO(28)xSU(2)\\ 
(0_4_)- SO(28)xSU(2)\\ 
(0_5_)- SO(24)xSU(2)xSU(2)\\ 
(0_6_)- SO(24)xSU(2)xSU(2)\\ 
(0_7_)- SO(24)xSU(2)xSU(2)\\ 
(3_4_)- SO(32)\\ 
(3_5_)- SO(28)xSU(2)\\ 
(3_7_)- SO(28)xSU(2)\\ 
(4_6_)- SO(28)xSU(2)\\ 
(4_7_)- SO(28)xSU(2)\\ 
(5_6_)- SO(24)xSU(4)\\ 
(5_7_)- SO(24)xSU(2)xSU(2)\\ 
(6_7_)- SO(24)xSU(2)xSU(2)\\ 
(0_3_4_)- SO(32)\\ 
(0_3_5_)- SO(28)xSU(2)\\ 
(0_3_7_)- SO(28)xSU(2)\\ 
(0_4_6_)- SO(28)xSU(2)\\ 
(0_4_7_)- SO(28)xSU(2)\\ 
(0_5_6_)- SO(24)xSU(4)\\ 
(0_5_7_)- SO(24)xSU(2)xSU(2)\\ 
(0_6_7_)- SO(24)xSU(2)xSU(2)\\ 
(3_4_7_)- SO(32)\\ 
(3_5_7_)- SO(28)xSU(2)\\ 
(4_6_7_)- SO(28)xSU(2)\\ 
(5_6_7_)- SO(24)xSO(8)\\ 
(0_3_4_7_)- SO(32)\\ 
(0_3_5_7_)- SO(28)xSU(2)\\ 
(0_4_6_7_)- SO(28)xSU(2)\\ 
(0_5_6_7_) = 0: SO(24)xSO(8)\\ 
\end{multicols}
\section{Vertices of the Polytopes presented in this paper}
\label{appendixD}
	$M0$: $
\left(\left(1,\,0,\,0\right), \left(0,\,1,\,0\right),
\left(0,\,0,\,1\right), \left(-1,\,-1,\,-1\right)\right)
$\\
$M2$: $
\left(\left(1,\,0,\,0\right), \left(0,\,1,\,0\right),
\left(0,\,0,\,1\right), \left(-3,\,-1,\,-1\right)\right)
$\\
$M3$: $
\left(\left(1,\,0,\,0\right), \left(0,\,1,\,0\right),
\left(-1,\,-1,\,0\right), \left(0,\,0,\,1\right),
\left(-1,\,0,\,-1\right)\right)
$\\
$M4$: $
\left(\left(1,\,0,\,0\right), \left(-1,\,0,\,0\right),
\left(0,\,1,\,0\right), \left(0,\,0,\,1\right),
\left(0,\,-1,\,-1\right)\right)
$\\
$M5$: $
\left(\left(1,\,0,\,0\right), \left(-1,\,0,\,0\right),
\left(0,\,1,\,0\right), \left(0,\,0,\,1\right),
\left(1,\,-1,\,-1\right)\right)
$\\
$M6$: $
\left(\left(1,\,0,\,0\right), \left(0,\,1,\,0\right),
\left(-1,\,-1,\,0\right), \left(0,\,0,\,1\right),
\left(1,\,0,\,-1\right)\right)
$\\
$M7$: $
\left(\left(1,\,0,\,0\right), \left(-1,\,0,\,0\right),
\left(0,\,1,\,0\right), \left(0,\,0,\,1\right),
\left(2,\,-1,\,-1\right)\right)
$\\
$M10$: $
\left(\left(1,\,0,\,0\right), \left(0,\,1,\,0\right),
\left(-2,\,-1,\,0\right), \left(0,\,0,\,1\right),
\left(-2,\,0,\,-1\right)\right)
$\\
$M11$: $
\left(\left(1,\,0,\,0\right), \left(0,\,1,\,0\right),
\left(-2,\,-1,\,0\right), \left(0,\,0,\,1\right),
\left(-1,\,1,\,-1\right)\right)
$\\
$M16$: $
\left(\left(1,\,0,\,0\right), \left(0,\,1,\,0\right),
\left(0,\,0,\,1\right), \left(-2,\,-1,\,-1\right),
\left(-1,\,1,\,0\right)\right)
$\\
$M88$: $
\left(\left(1,\,0,\,0\right), \left(0,\,1,\,0\right),
\left(0,\,0,\,1\right), \left(-6,\,-4,\,-1\right)\right)
$\\
$M221$: $
\left(\left(1,\,0,\,0\right), \left(0,\,1,\,0\right),
\left(0,\,0,\,1\right), \left(-5,\,-3,\,-1\right),
\left(-1,\,-1,\,1\right)\right)
$\\
$M230$: $
\left(\left(1,\,0,\,0\right), \left(0,\,1,\,0\right),
\left(1,\,-1,\,0\right), \left(0,\,0,\,1\right),
\left(-4,\,-2,\,-1\right)\right)
$\\
$M473$: $
\left(\left(1,\,0,\,0\right), \left(0,\,1,\,0\right),
\left(0,\,0,\,1\right), \left(-4,\,-3,\,-1\right),
\left(-1,\,0,\,1\right), \left(-2,\,-1,\,1\right)\right)
$\\
$M476$: $
\left(\left(1,\,0,\,0\right), \left(0,\,1,\,0\right),
\left(0,\,0,\,1\right), \left(-4,\,-2,\,-1\right),
\left(-5,\,-3,\,-1\right), \left(-1,\,-1,\,1\right)\right)
$\\
$M497$: $
\left(\left(1,\,0,\,0\right), \left(0,\,1,\,0\right),
\left(-1,\,1,\,0\right), \left(0,\,0,\,1\right),
\left(-2,\,-3,\,-1\right), \left(0,\,-1,\,1\right)\right)
$\\
$M859$: $
\left(\left(1,\,0,\,0\right), \left(0,\,1,\,0\right),
\left(1,\,-1,\,0\right), \left(0,\,0,\,1\right),
\left(-3,\,-1,\,-1\right), \left(0,\,-1,\,1\right),
\left(-1,\,-1,\,1\right)\right)
$\\
$M866$: $
\left(\left(1,\,0,\,0\right), \left(0,\,1,\,0\right),
\left(0,\,0,\,1\right), \left(-3,\,-2,\,-1\right),
\left(-1,\,0,\,1\right), \left(-4,\,-3,\,-1\right),
\left(-2,\,-1,\,1\right)\right)
$\\
$M895$: $
\left(\left(1,\,0,\,0\right), \left(0,\,1,\,0\right),
\left(-1,\,1,\,0\right), \left(0,\,0,\,1\right),
\left(-2,\,-2,\,-1\right), \left(-2,\,-3,\,-1\right),
\left(0,\,-1,\,1\right)\right)
$\\
$M1328$: $
\left(\left(1,\,0,\,0\right), \left(0,\,1,\,0\right),
\left(-1,\,1,\,0\right), \left(0,\,0,\,1\right),
\left(-1,\,-2,\,-1\right), \left(0,\,-1,\,1\right),
\left(-2,\,-2,\,-1\right),\right.
$\\
$\left. \left(-2,\,-3,\,-1\right)\right)$
\bibliographystyle{unsrt}
\bibliography{Bibli}

\begin{thebibliography}{10}

\bibitem{vafa_evidence_1996}
Cumrun Vafa.
\newblock Evidence for {{F}}-{{Theory}}.
\newblock {\em Nuclear Physics B}, 469(3):403--415, June 1996.

\bibitem{morrison_compactifications_1996-1}
David~R. Morrison and Cumrun Vafa.
\newblock Compactifications of {{F}}-{{Theory}} on {{Calabi}}--{{Yau
  Threefolds}} -- {{I}}.
\newblock {\em Nuclear Physics B}, 473(1-2):74--92, August 1996.

\bibitem{morrison_compactifications_1996}
David~R. Morrison and Cumrun Vafa.
\newblock Compactifications of {{F}}-{{Theory}} on {{Calabi}}--{{Yau
  Threefolds}} -- {{II}}.
\newblock {\em Nuclear Physics B}, 476(3):437--469, September 1996.

\bibitem{berglund_heterotic_1998}
P.~Berglund and P.~Mayr.
\newblock Heterotic {{String}}/{{F}}-theory {{Duality}} from {{Mirror
  Symmetry}}.
\newblock {\em arXiv:hep-th/9811217}, November 1998.

\bibitem{narain_new_1986}
K.~S. Narain.
\newblock New heterotic string theories in uncompactified dimensions {$<$} 10.
\newblock {\em Physics Letters B}, 169(1):41--46, March 1986.

\bibitem{aspinwall_k3_1996}
Paul~S. Aspinwall.
\newblock K3 {{Surfaces}} and {{String Duality}}.
\newblock {\em arXiv:hep-th/9611137}, November 1996.

\bibitem{schuett_elliptic_2009}
Matthias Schuett and Tetsuji Shioda.
\newblock Elliptic {{Surfaces}}.
\newblock {\em arXiv:0907.0298 [math]}, July 2009.

\bibitem{laza_arithmetic_2013}
Radu Laza, Matthias Schütt, and Noriko Yui, editors.
\newblock {\em Arithmetic and {{Geometry}} of {{K3 Surfaces}} and
  {{Calabi}}–{{Yau Threefolds}}}.
\newblock Fields {{Institute Communications}}. {Springer-Verlag}, {New York},
  2013.

\bibitem{laza_calabi-yau_2015-1}
Radu Laza, Matthias Schütt, and Noriko Yui, editors.
\newblock {\em Calabi-{{Yau Varieties}}: {{Arithmetic}}, {{Geometry}} and
  {{Physics}}: {{Lecture Notes}} on {{Concentrated Graduate Courses}}}.
\newblock Fields {{Institute Monographs}}. {Springer-Verlag}, {New York}, 2015.

\bibitem{kreuzer_classification_1998}
M.~Kreuzer and H.~Skarke.
\newblock Classification of {{Reflexive Polyhedra}} in {{Three Dimensions}}.
\newblock {\em arXiv:hep-th/9805190}, May 1998.

\bibitem{cardoso_duality_1996}
Gabriel~Lopes Cardoso, Gottfried Curio, Dieter Lust, and Thomas Mohaupt.
\newblock On the {{Duality}} between the {{Heterotic String}} and
  {{F}}-{{Theory}} in 8 {{Dimensions}}.
\newblock {\em Physics Letters B}, 389(3):479--484, December 1996.

\bibitem{candelas_f-theory_1997}
Philip Candelas and Harald Skarke.
\newblock F-theory, {{SO}}(32) and {{Toric Geometry}}.
\newblock {\em Physics Letters B}, 413(1-2):63--69, November 1997.

\bibitem{malmendier_k3_2015}
Andreas Malmendier and David~R. Morrison.
\newblock K3 surfaces, modular forms, and non-geometric heterotic
  compactifications.
\newblock {\em Letters in Mathematical Physics}, 105(8):1085--1118, August
  2015.

\bibitem{the_sage_developers_sagemath_2019}
\{The Sage Developers\}.
\newblock \{\vphantom\}{{S}}\vphantom\{\}{{ageMath}}, the
  \{\vphantom\}{{S}}\vphantom\{\}age \{\vphantom\}{{M}}\vphantom\{\}athematics
  \{\vphantom\}{{S}}\vphantom\{\}oftware \{\vphantom\}{{S}}\vphantom\{\}ystem
  (\{\vphantom\}{{V}}\vphantom\{\}ersion 8.6), 2019.

\bibitem{braun_palp_2012}
Andreas~P. Braun, Johanna Knapp, Emanuel Scheidegger, Harald Skarke, and
  Nils-Ole Walliser.
\newblock {{PALP}} - a {{User Manual}}.
\newblock {\em arXiv:1205.4147 [hep-th]}, pages 461--550, December 2012.

\bibitem{kreuzer_palp_2004}
Maximilian Kreuzer and Harald Skarke.
\newblock {{PALP}}: {{A Package}} for {{Analyzing Lattice Polytopes}} with
  {{Applications}} to {{Toric Geometry}}.
\newblock {\em Computer Physics Communications}, 157(1):87--106, February 2004.

\bibitem{batyrev_dual_1993}
Victor~V. Batyrev.
\newblock Dual {{Polyhedra}} and {{Mirror Symmetry}} for {{Calabi}}-{{Yau
  Hypersurfaces}} in {{Toric Varieties}}.
\newblock {\em arXiv:alg-geom/9310003}, October 1993.

\bibitem{cox_toric_nodate}
David~A. Cox, John~B. Little, and Henry~K. Schenck.
\newblock {\em Toric Varieties}.
\newblock {American Mathematical Society}, c2011.

\bibitem{skarke_string_1999}
Harald Skarke.
\newblock String {{Dualities}} and {{Toric Geometry}}: {{An Introduction}}.
\newblock {\em Chaos, Solitons \& Fractals}, 10(2-3):543--554, February 1999.

\bibitem{candelas_duality_1998}
Philip Candelas and Anamaria Font.
\newblock Duality {{Between}} the {{Webs}} of {{Heterotic}} and {{Type II
  Vacua}}.
\newblock {\em Nuclear Physics B}, 511(1-2):295--325, February 1998.

\bibitem{kodaira_compact_1963}
K.~Kodaira.
\newblock On {{Compact Analytic Surfaces}}: {{II}}.
\newblock {\em Annals of Mathematics}, 77(3):563--626, 1963.

\bibitem{neron_modeles_1964}
André Néron.
\newblock {Modèles minimaux des variétés abéliennes sur les corps locaux et
  globaux}.
\newblock {\em Publications Mathématiques de l'IHÉS}, 21:5--128, 1964.

\bibitem{perevalov_enhanced_1997}
Eugene Perevalov and Harald Skarke.
\newblock Enhanced gauge symmetry in type {{II}} and {{F}}-theory
  compactifications: {{Dynkin}} diagrams from polyhedra.
\newblock {\em Nuclear Physics B}, 505(3):679--700, November 1997.

\bibitem{mayrhofer_mordell-weil_2014}
Christoph Mayrhofer, David~R. Morrison, Oskar Till, and Timo Weigand.
\newblock Mordell-{{Weil Torsion}} and the {{Global Structure}} of {{Gauge
  Groups}} in {{F}}-theory.
\newblock {\em Journal of High Energy Physics}, 2014(10):16, October 2014.

\bibitem{braun_geometric_2013}
Volker Braun, Thomas~W. Grimm, and Jan Keitel.
\newblock Geometric {{Engineering}} in {{Toric F}}-{{Theory}} and {{GUTs}} with
  {{U}}(1) {{Gauge Factors}}.
\newblock {\em Journal of High Energy Physics}, 2013(12), December 2013.

\bibitem{klevers_f-theory_2015}
Denis Klevers, Damian Kaloni~Mayorga Pena, Paul-Konstantin Oehlmann, Hernan
  Piragua, and Jonas Reuter.
\newblock F-{{Theory}} on all {{Toric Hypersurface Fibrations}} and its {{Higgs
  Branches}}.
\newblock {\em J. High Energ. Phys.}, 2015(1):142, January 2015.

\bibitem{cvetic_discrete_2017}
Mirjam Cvetic, Antonella Grassi, and Maximilian Poretschkin.
\newblock Discrete {{Symmetries}} in {{Heterotic}}/{{F}}-theory {{Duality}} and
  {{Mirror Symmetry}}.
\newblock {\em J. High Energ. Phys.}, 2017(6):156, June 2017.

\bibitem{cvetic_tasi_2018}
Mirjam Cvetic and Ling Lin.
\newblock {{TASI Lectures}} on {{Abelian}} and {{Discrete Symmetries}} in
  {{F}}-theory.
\newblock {\em arXiv:1809.00012 [hep-th]}, August 2018.

\bibitem{grassi_weierstrass_2012}
Antonella Grassi and Vittorio Perduca.
\newblock Weierstrass models of elliptic toric {{K3}} hypersurfaces and
  symplectic cuts.
\newblock {\em arXiv:1201.0930 [hep-th]}, January 2012.

\bibitem{candelas_type_2015}
Philip Candelas, Andrei Constantin, Cesar Damian, Magdalena Larfors, and
  Jose~Francisco Morales.
\newblock Type {{IIB}} flux vacua from {{G}}-theory {{I}}.
\newblock {\em Journal of High Energy Physics}, 2015(2), February 2015.

\bibitem{berglund_periods_1994}
Per Berglund, Philip Candelas, Xenia {de la Ossa}, Anamaria Font, Tristan
  Hubsch, Dubravka Jancic, and Fernando Quevedo.
\newblock Periods for {{Calabi}}--{{Yau}} and {{Landau}}--{{Ginzburg Vacua}}.
\newblock {\em Nuclear Physics B}, 419(2):352--403, May 1994.

\bibitem{font_comments_nodate-1}
Anamaría Font, Christoph Mayrhofer, and Hector Parra.
\newblock Comments on {{F}}-theory/heterotic duality in 8 dimensions.
\newblock To appear.

\bibitem{font_comments_2019}
Anamaría Font.
\newblock Comments on {{F}}-theory/heterotic duality in 8 dimensions, June
  2019.
\newblock
  indico.cern.ch/event/782271/contributions/3439041/attachments/1865076/3066391/cs4.pdf.

\bibitem{blumenhagen_basic_2013}
Ralph Blumenhagen, Dieter Lüst, and Stefan Theisen.
\newblock {\em Basic {{Concepts}} of {{String Theory}}}.
\newblock Theoretical and {{Mathematical Physics}}. {Springer Berlin
  Heidelberg}, {Berlin, Heidelberg}, 2013.

\bibitem{fraiman_new_2018}
Bernardo Fraiman, Mariana Graña, and Carmen~A. Núñez.
\newblock A new twist on heterotic string compactifications.
\newblock {\em Journal of High Energy Physics}, 2018(9), September 2018.

\end{thebibliography}
\end{document}